\documentclass[]{acmart}
\usepackage{hyperref}
\usepackage{subcaption}
\usepackage{pifont}
\usepackage{enumitem}

\newcommand{\reftable}[1]{\textbf{Table~\ref{#1}}}
\newcommand{\refsec}[1]{\textbf{Section~\ref{#1}}}
\newcommand{\reffig}[1]{\textbf{Fig.~\ref{#1}}}

\newcommand{\githubrepo}{\url{https://github.com/onspatial/wastewater-based-epidemiology-patterns-of-life}}
\newcommand{\fullresultsnote}{Full results for infection rates 0.1, 0.125, 0.15, 0.175, 0.2, up to 0.5 are available on GitHub.}

\newcommand{\andreas}[1]{}
\newcommand{\andrew}[1]{}

\newcommand{\cmark}{\ding{51}}
\newcommand{\xmark}{\ding{55}}

\renewcommand\footnotetextcopyrightpermission[1]{}

\begin{document}

%\title{Wastewater-Based Epidemiology Using Patterns of Life: An Agent-Based Simulation for Linking Human Behavior to Pathogen Dynamics}
\title{Where do We Poop? City-Wide Simulation of Defecation Behavior for Wastewater-Based Epidemiology}
% Where Poop Spread the Pathogen: An Agent-based geosimulation for Wastewater Epidemiology
\author{Hossein Amiri}
\orcid{0000-0003-0926-7679}
\email{hossein.amiri@emory.edu}
\affiliation{%
    \institution{Emory University}
    \city{Atlanta}
    \country{USA}
}

\author{Akshay Deverakonda}
\orcid{0009-0008-9899-271X}
\email{Akshay.deverakonda@emory.edu}
\affiliation{%
    \institution{Emory University}
    \city{Atlanta}
    \country{USA}
}

\author{Yuke Wang}
\orcid{0000-0002-9615-7859}
\email{yuke.wang@emory.edu}
\affiliation{%
    \institution{Emory University}
    \city{Atlanta}
    \country{USA}
}

\author{Andreas Z{\"u}fle}
\orcid{0000-0001-7001-4123}
\email{azufle@emory.edu}
\affiliation{%
    \institution{Emory University}
    \city{Atlanta}
    \country{USA}
}

\renewcommand{\shortauthors}{Amiri, et al.}
\begin{abstract}

Wastewater surveillance, which regularly measures pathogen biomarkers in wastewater samples, is a valuable tool for monitoring infectious diseases circulating in communities. Yet, most wastewater-based epidemiology methods that use wastewater surveillance results to infer disease trends implicitly assume that individuals excrete only at their residential locations and that the populations contributing to wastewater samples are static. These simplifying assumptions ignore daily mobility, social interactions, and heterogeneous toilet-use patterns, which can bias the interpretation of wastewater results, especially at upstream sampling locations such as neighborhoods, institutions, or buildings. Here, we introduce an agent-based geospatial simulation framework. Building on an established Patterns of Life model, we simulate daily human activities within a realistic urban environment and extend the framework with a physiologically motivated defecation cycle and toilet-use patterns. We couple this behavioral model with an infectious disease model to simulate transmission through spatial and social interactions. When an infected agent defecates, a pathogen-shedding model determines the amount of pathogen released in the feces. By integrating population mobility, disease transmission, toilet-use behavior, and pathogen shedding, the framework can simulate the spatiotemporal dynamics of wastewater pathogen loads. Using a case study of 10,000 simulated agents in Fulton County, Georgia, we examine how varying infection rates alter epidemic trajectories, wastewater pathogen loads, and the spatial distribution of pathogen shedding over time. Our results show that mobility and toilet use can substantially decouple residential disease prevalence from wastewater pathogen loads and demonstrate how behaviorally grounded simulations can support interpretation, scenario analysis, and wastewater surveillance strategies designs.
   
\end{abstract}


\maketitle
\pagestyle{plain}

\section{Introduction}
\label{sec:introduction}

Wastewater surveillance (WWS) is a public health tool that regularly measures pathogens in wastewater samples, which are pooled samples of feces, urine, and sputum from the population using sewage systems, to monitor infectious disease prevalence in the community. Since the start of the COVID-19 pandemic, this novel approach has been widely implemented to supplement the use of epidemiological case surveillance for rapidly identifying disease outbreaks \cite{scott2021targeted, wang2022early}, monitoring temporal and spatial trends in disease transmission \cite{hopkins2023citywide, wang2023case}, and guiding disease prevention and control measures \cite{thompson2020making, hopkins2023public}.  
\begin{figure}
    \centering
    \includegraphics[width=\linewidth]{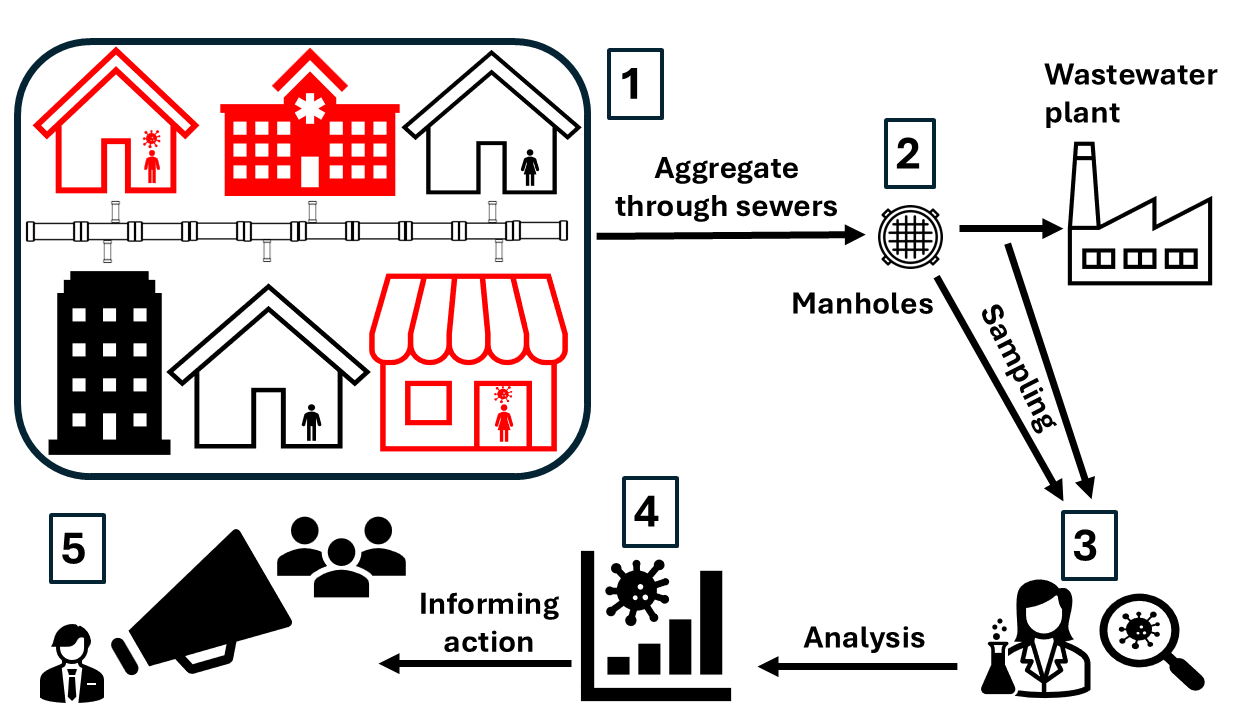}
    \caption{Conceptual overview of wastewater surveillance. (1) In a specific geographic area covered by a sewer system, referred to as a sewershed, infected (in red) and non-infected (in black) persons from different buildings contribute fecal waste into a sewer system. (2) We can collect wastewater samples either from manholes located across the sewer system or at the downstream wastewater treatment plant (influent line). (3) We can analyze samples to detect the presence of pathogens and quantify their levels. (4 and 5) The WWS results can help us estimate trends in pathogen prevalence over time in the sewershed, informing public health response. \newline {\it \footnotesize *Sewer manhole icon used in this diagram obtained from Flaticon.com.}}
    \Description{}
    \label{fig:pathogen_in_system}
\end{figure}

\reffig{fig:pathogen_in_system} illustrates the WWS process: (1) Assume we have a study region, such as a city, in which individuals carrying an infectious diseases live. In \reffig{fig:pathogen_in_system}, we highlight homes and a hospital housing infectious agents in red color. People use the toilet and, depending on their disease status, release loads of pathogens into their local sewer system. The pathogens travel down-stream from local wastewater pipes of individual homes into the sewer network of the local neighborhood, into larger sewer network of increasingly larger neighborhoods until eventually reaching wastewater treatment plants (WWTPs) where wastewater is collected. (2) WWS collects wastewater samples at any point in the sewer network: At the WWTP or at any point upstream from sewer manholes which may collect wastewater from communities, neighborhoods, or individual homes. (3) The collected samples are analyzed in labs to detect the load of a pathogen using highly sensitive testing. For example, for COVID-19, pathogens shed by as few as 5-10 actively shedding individuals can be detected with high sensitivity among a population of 100,000.

As few as one infected person among tens of thousands can still be detected~\cite{hewitt2022sensitivity}. (4) These wastewater results are analyzed for spatial and temporal trends, to identify local outbreaks, such as outbreaks at student residence halls at Emory University~\cite{wang2022early}. (5) Finally, these trends are communicated to policymakers such as leaders in university administration or state, indigenous, local, or territorial health departments to take preventive action such as informing residents, preemptive resource allocation, and protecting vulnerable communities. 

Whereas WWS centers on detecting pathogens and tracking their concentrations across space and time, wastewater-based epidemiology (WBE) goes a step further, drawing on these measurements to infer population-level epidemiological trends, both spatial and temporal. For simplicity, most WBE studies assume that individuals only use toilets in their residential location~\cite{medema2020implementation, sims2020future} without considering human mobility. 
Numerous recent WBE programs have been used for sentinel surveillance to monitor specific sites in dense urban areas  (e.g., hospitals, university campuses, airports, conference centers, stadiums, and correctional facilities) \cite{gibas2021implementing, wang2022early, farkas2023wastewater, nkambule2024wastewater, saber2024correlation}. Our hypothesis is that ignoring human mobility may underestimate disease prevalence in residential areas while overestimating it in commercial areas. For example, a commercial area having many workplaces but few residential houses would attribute all pathogens from many workers to a few residents. For instance, during the COVID-19 pandemic, shelter-in-place orders, which required all residents and visitors to remain in their residences and limit social interactions, led to a decline in SARS-CoV-2 concentrations in wastewater samples from sites capturing commercial areas. This decrease could reflect reduced population mobility rather than a decline in disease burden, highlighting how biased estimations can complicate disease trend analysis and public health decision-making~\cite{thomas2017use}. 
% \andreas{This is a bold statement that existing research is flawed. This statement needs to be supported with reference and/or a strong justification}. 
Thus, we postulate that it is critical to consider human mobility rather than assuming home-only toilet use. %Meanwhile, this assumption makes WBE unable to accurately interpret the changes in wastewater results. 
%A sewershed refers to the geographic area from which wastewater flows into a specific point (in this case, sampling point of WWS) in the sewer system. 
Towards this, the goal of this study is to drop the assumption of only using toilets at home by using geosimulation to (1) simulate realistic spatial pooping behavior at the individual level, (2) simulate an infectious disease outbreak within the simulated population and the corresponding release of pathogens (shedding) into the sewer network, and (3) share the simulation framework and generated example datasets to support WBE.

%When a WWS sampling downstream (city level) such as at wastewater treatment facility, a majority of toilet use of residents may be still occur within the same sewershed lead to less biased estimation. But this bias increases as WWS sampling sites moving upstream to community level, institutional level, and building level \cite{sims2020future}. 

Recent studies have begun to integrate agent-based infection dynamics, human mobility, pathogen shedding, and wastewater transport within unified frameworks for WBE \cite{schmid2025integrative,0616456898e04215a978dfebbfd6f6f5}.
 However, to the best of our knowledge, no existing model explicitly combines high-resolution individual-level population mobility, social contact patterns, toilet-use behavior (defecation events and toilet choice), and mechanistic  shedding dynamics in a single scalable agent-based geospatial simulation that is capable to generate the number of pathogens enters the sewer system through a toilet at a specific time. Connecting this model with models of pathogen fate in sewerage, WWS, and public health interventions will create a comprehensive ``playground'' for conducting scenario studies that assess how public health interventions altering human behavior can reduce disease transmissions. 
 In this paper, we address this gap by proposing an agent-based geospatial simulation that explicitly links human behavior, infectious disease dynamics, and pathogen load in wastewater. With these insights, our main contributions are as follows:
 
\begin{itemize}[noitemsep, leftmargin=15pt, labelsep=7pt]

    \item We develop a scalable agent-based geospatial simulation framework that links individual human behavior end-to-end to the sewer system, producing time- and location-resolved pathogen inputs (i.e., how many pathogens enter the sewer through a given toilet at a given time).  The framework integrates an SEIR infection model with multilayer transmission via co-location and social networks, heterogeneous disease progression times, and a gamma-like pathogen-shedding function that connects transmission dynamics to fecal pathogen loads in wastewater.
    \item We demonstrate the framework in a case study of 10,000 agents in Fulton County, Georgia, USA, and show how mobility patterns, infection rates, and toilet use jointly shape epidemic curves, the spatial spread of infection, and the temporal dynamics of pathogen loads. We further show how human mobility affects shedding events and shedd pathogen levels using a specific sewershed within Fulton County.
    \item We release the source code as an open-source project on GitHub:\\ \githubrepo
\end{itemize}

The remainder of the paper is organized as follows. 
\refsec{sec:background} provides background on the WWS, WBE, and Patterns of Life simulation. 
\refsec{sec:related_works} reviews related work on dynamic populations in sewersheds, mobility informed wastewater analysis, and simulation based approaches. 
\refsec{sec:methodology} presents our methodology, including the extended behavioral dynamics, infectious disease and shedding models, and parameter settings. 
\refsec{sec:results} reports experimental results for the Fulton County case study, analyzing disease dynamics, pathogen loads, and spatial patterns of spread. 
\refsec{sec:discussions} discusses implications for interpreting pathogen loads in wastewater and designing surveillance strategies. 
\refsec{sec:conclusion} concludes and outlines directions for future research.

\section{Background}
\label{sec:background}
In this section, we provide the background information, including an overview of WWS, WBE, and the Patterns of Life simulation. 

%\andrew{TO DISCUSS: Do you think we can have a table for those metrics you are tracking for each agent? And how frequent...}
\subsection{Wastewater surveillance}
WWS is a public health monitoring approach that analyzes sewage samples to detect and track diseases or other health-related issues within a community. Unlike traditional epidemiological surveillance, which conducts individual-level clinical testing, WWS examines what entire populations collectively excrete into the sewer system, providing a community-level snapshot of health status. The concept has roots dating back to the mid-20th century when researchers successfully used wastewater to monitor poliovirus circulation in different geographic regions, demonstrating the feasibility of tracking infectious diseases through sewage systems~\cite{paul1941virus, trask1942periodic, rhodes1950poliomyelitis}.

The WWS process involves collecting wastewater from strategic sampling sites in the sewerage system--typically at downstream WWTPs or specific upstream locations--and analyzing these samples in the laboratory to detect genetic material or other biomarkers of interest. The resulting data can reveal patterns of disease presence and spread across communities served by the sewer system. This approach offers several compelling advantages for public health agencies. It provides population-level monitoring without requiring individual testing~\cite{thompson2020making}, can detect signals from both symptomatic and asymptomatic infections~\cite{medema2020implementation}, and often serves as an early warning system for disease outbreaks~\cite{mao2020potential, wang2022early}. Additionally, WWS is generally more cost-effective than mass individual testing programs and can capture information from under-served populations that may not have adequate testing capacity.

\subsection{Wastewater-based epidemiology}
WBE, building upon the infrastructure of WWS, establishes a comprehensive framework to infer health outcomes from wastewater measurements. While WWS focuses primarily on detecting and monitoring the presence of targeted biomarkers in wastewater, WBE represents a broader analytical approach that uses wastewater measurements to draw inferences about population health, behaviors, and exposures~\cite{sims2020future, polo2020making}. Essentially, WBE treats wastewater as a pooled biological sample that captures the collective biochemical ``fingerprint'' of a community.

During the COVID-19 pandemic, WBE gained prominence as research institutions and organizations and government agencies worldwide implemented WWS for SARS-CoV-2, demonstrating its value for tracking pandemic trends, identifying emerging hotspots, and providing early warnings of surges in infection at localized scales such as university campuses and residential facilities~\cite{scott2021targeted, wang2022early, wang2023case, hopkins2023citywide}. The scope of WBE encompasses diverse applications beyond infectious disease monitoring. Researchers have used this approach to track pharmaceutical consumption, estimate illicit drug use, assess nutritional status through metabolic markers, and monitor environmental chemical exposures at the population level~\cite{ort2010sampling, been2014population}.

The primary strength of WBE lies in its objectivity and near-real-time nature, as it captures unbiased population-level data independent of healthcare-seeking behavior or testing availability. It can reveal trends in hard-to-reach populations and detect asymptomatic infections that would otherwise go unnoticed in epidemiological surveillance systems~\cite{medema2020implementation}. However, WBE also has important limitations. The data represent aggregated community-level information rather than individual cases, making it impossible to identify specific infected persons or their precise locations within a sewer catchment area~\cite{wang2023case}. Additionally, the relationship between wastewater pathogen loads and actual disease prevalence can be influenced by factors such as pathogen shedding variability among infected individuals, wastewater dilution, and environmental degradation of target biomarkers, as well as toilet-use behavior and population mobility~\cite{medema2020implementation, sims2020future, wade2022understanding}.

\subsection{Patterns of Life Simulation}
The pattern-of-life simulation~\cite{zufle2023urban} used in this study is an established agent-based framework that captures a broad range of human behaviors, including working, staying at home, sleeping, eating, and interacting with friends and strangers in public spaces such as restaurants and recreational sites. The framework is theory-driven, with behavioral logic grounded in established models of human activity. In particular, agents make decisions based on their needs, structured around Maslow’s hierarchy of needs~\cite{maslow1943theory}, encompassing physiological requirements (food, shelter, sleep), safety needs (financial safety), and social needs (love). Meeting these needs enables agents to sustain homeostasis and overall well-being. The design and validation of this framework have been thoroughly documented in prior studies~\cite{kim2020location,amiri2024patterns}. The simulation has been used as a baseline framework for data generation and has been extended with additional features, including anomalous agents and infectious disease modeling~\cite{amiri2023massive,kohn2023epipol,zhang2024large,zhang2024transferable,amiri2024urban}.

\subsubsection{Synthetic Population Generation}
The POL model differentiates behavioral patterns through heterogeneous agent attributes and need-driven decision making. Each agent is initialized with individual characteristics such as age, education, financial resources, movement speed, food need, sleep need, social need, home location, workplace or school assignment, and social relationships. These attributes affect both the priority of needs and the feasible actions available to the agent. At each simulation step, agents evaluate competing needs, including food, shelter, sleep, financial safety, and social connection, and select actions that satisfy the most pressing needs subject to constraints such as time of day, work schedule, money, distance, venue capacity, and spatial accessibility.

\subsubsection{Emerging Patterns of Life}
As a result of different attributes and needs, behavioral patterns differ across agents: for example, agents with different jobs follow different work schedules, agents with different food needs visit restaurants at different times, agents with different social needs participate in different levels of social activity, and agents living or working in different neighborhoods generate distinct mobility trajectories. This design creates emergent patterns of life unique to each agent to satisfy their needs rather than having agents choose random destinations.

\subsubsection{OpenStreetMap-Based Environment}
The Patterns of Life simulation represents everyday human activity within real geographic environments using OpenStreetMap (OSM) data, including roads and buildings. OSM tags are used to define residential buildings and commercial building types (restaurants, recreational sites, workplaces).  

\subsubsection{Simulation of Time}
Time progresses in discrete steps mapped to real clock time, allowing the model to track hours, days, weekdays, and weekends, as well as calendar events such as national holidays, while agents live in homes, work, eat out, and socialize. 
The simulation explicitly models daily rhythms. At midnight, global updates process rent, tuition, aging, and financial balances and agents generate plans for the following day. %Nightly summaries compute venue visitation profiles, apply decay to social ties, prune weak relationships, and calculate the expected strength of stable connections. 
Agents follow daily plans balancing home, work, meals, and recreation based on their needs. 

\subsubsection{Social Networks}
Two directed social graphs are maintained: one representing family and friendship networks, and another capturing work relationships. Social networks emerge dynamically through co-location: interactions strengthen ties, whereas inactivity causes them to decay. To satisfy their social need, which is measured by their number of links in the social networks, agents make plans to meet existing friends or to make new friends.

\section{Related Works}
\label{sec:related_works}

% In \cite{} x is done and the results is y

\subsection{Measures of Dynamic Populations in Sewersheds}

In wastewater surveillance, population normalization enables comparisons of wastewater measurements from the same sampling location at different times and across sampling locations. However, estimates of the catchment population used for normalization have varying degrees of uncertainty depending on the methodology applied \cite{boogaerts2024current}.

Two common methodologies are used to normalize catchment populations. The first uses census data, which are generally accessible and inexpensive, if not free, to obtain \cite{price2024testing}. A study in France examined the relationship between wastewater flow rate and mobility into and out of a catchment area, as determined from census surveys \cite{atinkpahoun2018population}. The study found that changes in mobility correlated with changes in certain pollutants over the study period. Another study used census data to estimate catchment populations and found that the estimates differed from population numbers reported by WWTPs \cite{tscharke2019harnessing}. Census data also provide sociodemographic characteristics, such as income and age, that can be associated with a catchment area \cite{boogaerts2024current}. However, population-normalization methods based on census data are not consistently reported in the literature, and different methods can produce substantially different estimates \cite{price2024testing}. In addition, wastewater catchment boundaries may not align with census tracts \cite{boogaerts2024current}.

\subsection{Mobility and Wastewater Data}
One approach to estimating dynamic populations in wastewater surveillance uses mobile-device or other signaling data. Compared with biomarkers, mobile-device data can provide information about the near-real-time movement of people carrying phones \cite{baz2019assessing,thomas2017use}. Thomas et al. (2017) were among the first to use mobile-device data to show that the population within a catchment varied over timescales ranging from within a day to across a month \cite{thomas2017use}. They used this variability to calculate population-normalized loads of illicit drugs that accounted for dynamic populations over the study period. Another study used mobile-device data to improve understanding of psychoactive pharmaceutical use in a catchment over two years \cite{boogaerts2023lockdown}. Integrating mobility data improved interpretation of pharmaceutical-use trends by clarifying how the underlying population changed over time.

In addition to examining population-normalized loads of substances and pollutants, researchers have used mobility data to characterize wastewater flows. One study assessed correlations between mobility data and wastewater flows across five catchments in Sweden, with promising results that warrant further investigation using more refined modeling techniques \cite{neuenhofer2025exploring}.

A drawback of mobile-device data is that they can be expensive and require specialized expertise. This has motivated the search for less expensive proxies. One study found that ammonium correlates with mobility and can serve as a proxy for normalizing population changes in a catchment \cite{baz2019assessing}. Nevertheless, the value of mobility data remains substantial. Another study compared daily methamphetamine loads normalized using mobile-device data, total nitrogen, total phosphorus, biological oxygen demand, and census data. Mobile-device data performed substantially better than the other sources for estimating real-time population changes, especially within a specific catchment rather than across an entire metropolitan area \cite{sim2023evaluation}.

\subsection{Simulating Dynamic Populations in Wastewater Sewersheds}
Although the literature includes many methods for characterizing dynamic populations in sewersheds, most focus on changes in the total catchment population over time. Fewer studies examine how dynamic populations vary across space and how that spatial distribution changes over time. One previous study developed an agent-based model to simulate wastewater production, particularly from infected agents during a simulated disease outbreak \cite{0616456898e04215a978dfebbfd6f6f5}.

\section{Methodology}
\label{sec:methodology}
% \textcolor{red}{We could have a diagram to show how a day of infected agent to highlight key features (moving, eating, pooping) for this model. It will be good to demonstrate how the pooping occurs (a combination of pooping need and conditions, toilet available). It will be important for the public health audiences.}

In this section, we describe the simulation dynamics that capture daily and weekly human behavioral patterns at five-minute resolution, the infectious disease framework governing transmission and recovery, the mechanistic representation of defecation events and pathogen shedding into wastewater, and the analytical methods used to process and interpret the resulting data.

\begin{table}
    \centering
    \caption{Behavioral and physiological needs represented in the extended Patterns of Life simulation compared with the baseline version.}
    \label{tab:old-new-comparision}
    \begin{tabular}{lccccccc}
         & Food  & Love  & Shelter    & Sleep  & Financial Safety  & Defecation  & Infectious Disease \\
        \hline
        Extended Version& \cmark  & \cmark & \cmark & \cmark & \cmark & \cmark & \cmark \\
        Baseline Version & \cmark  & \cmark & \cmark & \cmark & \cmark & \xmark & \xmark \\
    \end{tabular}
\end{table}

\subsection{Simulation of Defecation Behavior}
The Patterns of Life Simulation~\cite{zufle2023urban,amiri2024patterns} is an agent-based framework grounded in physiological and sociological mechanisms inspired by Maslow’s theory of human needs~\cite{maslow1943theory}. In this framework, an agent’s behavior is guided by its underlying needs, which drive decision making and daily activity patterns. The baseline model includes essential needs such as food, love, sleep, shelter, and financial safety, which together drive individual activities and social interactions. Following the same design principle, we introduce an additional physiological element to represent the defecation need, allowing agents to exhibit more natural and biologically consistent behavior. A comparison of the baseline model and the extended model presented in this paper is provided in \reftable{tab:old-new-comparision}. As shown, our updated framework introduces two additional components—defecation and infectious disease dynamics—that were not included in the vanilla version. These additions allow the simulation to capture more realistic behavioral and wastewater-based epidemiology patterns observed in populations.

\begin{figure}
    \centering
    \includegraphics[width=0.5\linewidth]{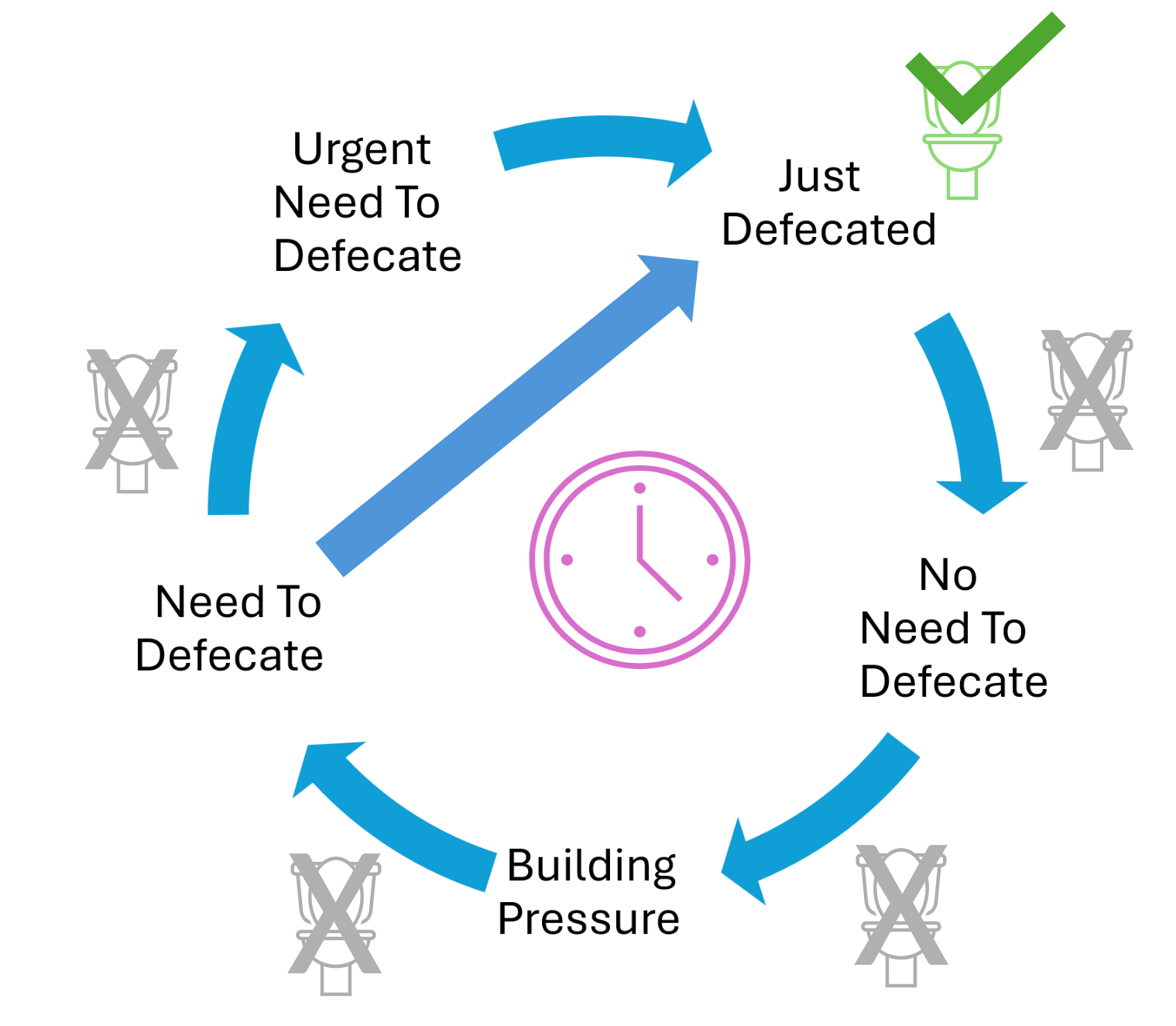}
    \caption{Defecation cycle within the simulation. The process begins with the \textit{`Just Defecated'} state and progresses through subsequent stages before returning to the start. The full cycle may repeat multiple times per day, depending on each agent’s individual defecation rate.}
    \Description{Five-state cycle from just defecated through no need, building pressure, need, and urgent need before returning to the initial state.}
    \label{fig:defecation_cycle}
\end{figure}

In the extended wastewater simulation, heterogeneity is also introduced in the physiological, social, and health components. Eating behavior and defecation behavior are modeled as separate agent-level processes rather than assuming a fixed one-to-one relationship. Each agent receives an individualized defecation rate, which controls how quickly the agent moves from a comfortable state to building pressure, need to defecate, and urgent need to defecate. This allows agents to differ in both the number and timing of defecation events during the day. Similarly, social dynamics vary through each agent’s social need and evolving friendship, family, and workplace networks, which affect where and with whom agents interact. Health dynamics are also heterogeneous: agents are assigned individual susceptibility, contagiousness, infection progression timing, and shedding-rate parameters. Therefore, two agents with the same residential neighborhood may still differ substantially in mobility, eating behavior, toilet-use timing, social exposure, infection risk, and pathogen contribution to wastewater. These agent-level differences generate the population-level variation in wastewater signals studied in this paper. We introduce a new physiological mechanism that captures the gradual buildup and relief of a biological need over time. Each agent continuously tracks its internal state, as shown in \reffig{fig:defecation_cycle}, which evolves based on elapsed time and an individual characteristic called the defecation rate, drawn from a uniform distribution. The process progresses through multiple stages, ranging from complete comfort to increasing urgency, depending on how much time has passed since the last relief event. These transitions are governed by specific thresholds that determine when an agent experiences mild, moderate, or urgent pressure. Once the need is satisfied, the internal state resets, initiating a new cycle. This continuous time-dependent regulation strengthens the model’s temporal consistency and introduces a natural rhythm into agent behavior.

After each relief event, the agent experiences a comfortable period that gradually transitions into rising internal pressure as simulated time passes. The rate of this buildup varies among agents, reflecting individual-level heterogeneity in biological rhythms and daily routines. 
When the internal pressure surpasses a defined threshold, the agent becomes increasingly aware of the need and eventually must take action to relieve it. The agent will only take this action when not in transit; for instance, while at home, work, or in a public location. These transitions are smooth and progressive rather than abrupt, enabling the simulation to represent realistic behavioral variability. 
Integrating this physiological mechanism within the broader hierarchy of needs allows the simulation to represent a more holistic view of human behavior. As internal pressure increases, agents may alter their plans or temporarily re-prioritize tasks to address their bodily needs, just as people do in real life. This dynamic interplay between biological, social, and environmental drivers generates emergent behavioral patterns that evolve naturally across the population. Differences in individual timing and urgency create realistic variations in activity peaks, movement density, and location demand.

More specifically, \reffig{fig:defecation_cycle} illustrates a five-state progression: \textit{`Just Defecated'} → \textit{`No Need To Defecate'} → \textit{`Building Pressure'} → \textit{`Need To Defecate'} → \textit{`Urgent Need To Defecate'}. At initialization, each agent is assigned a defecation rate sampled from a uniform distribution to capture inter-individual variability. This rate sets three timing controls: how long it takes to leave \textit{`No Need To Defecate'}, how long that comfort is maintained, and how quickly internal pressure builds per simulation step. A need threshold marks the transition into \textit{`Need To Defecate'}, and an urgent threshold marks \textit{`Urgent Need To Defecate'}. Each completed event increments a counter of defecations, and the internal state is reset, starting a new cycle.
At each step, the update logic advances the state based on elapsed minutes since the last event. \textit{`Just Defecated'} immediately becomes \textit{`No Need To Defecate'} with the comfort level restored. While in \textit{`No Need To Defecate'}, the agent remains comfortable until the combined reach and keep times pass, then enters \textit{`Building Pressure'}. In \textit{`Building Pressure'}, the comfort level decreases each step; crossing the need threshold moves the agent to \textit{`Need To Defecate'}. There, comfort continues to decline and is clamped at the urgent threshold; once it reaches that point, the state becomes \textit{`Urgent Need To Defecate'}. During the need-satisfaction step, if the agent is not in transit and is in \textit{`Need To Defecate'} or \textit{`Urgent Need To Defecate'}, the agent acts to relieve the need, returning to \textit{`Just Defecated'} and restarting the cycle. 

In addition, we separated the eating rate from the defecation rate to better reflect realistic physiological variation. Some individuals may eat large amounts but defecate only once, while others eat heavily and defecate frequently. Likewise, some people eat very little yet still defecate multiple times, and others eat little and rarely defecate. To capture this diversity, each agent is assigned a defecation rate $r$, drawn from a uniform distribution. This rate determines the timing of the entire physiological cycle. After defecation, an agent experiences a period with no urge to defecate lasting $60 - 30r$ minutes, followed by a comfortable period of $180 - 60r$ minutes during which internal pressure gradually increases. As time progresses, the agent's internal emptiness decreases at a rate proportional to $0.65r$. When this internal value drops below the threshold $30 + 20r$, the agent begins to feel the need to defecate, and when the value reaches zero the agent enters an urgent state. After defecation, the internal state resets and the cycle starts again.

\begin{table}[t]
\centering
\caption{Summary of simulation parameters governing defecation, disease progression, and pathogen transmission. Experimental overrides are reported in \reftable{tab:simulation_configuration}.}
\label{tab:sim-params}
\begin{tabular}{p{3cm} p{2.1cm} p{9cm}}
\hline
\textbf{Parameter} & \textbf{Default Value} & \textbf{Description} \\
\hline
Defecation rate & $r \sim U(0.2, 0.8)$ & Agent-specific rate governing the timing of the defecation cycle. \\
Incubation period & $D_E = 3$ days & Duration an exposed agent waits before becoming infectious. \\
Infectious period & $D_I = 3$ days & Number of days an infectious agent sheds pathogens. \\
Recovery period & $D_R = 3$ days & Time until recovered agents return to susceptible state. \\
Infection ratio & $p_{\text{inf}} = 0.1$ & Probability that a contact results in infection. \\
Spreading per person & $n_{\text{spread}} = 1$ & Maximum number of infection attempts by an infectious individual in one step. \\
Spreading per unit & $n_{\text{unit}} = 1$ & Maximum number of infection attempts associated with a shared spatial unit per agent. \\
Initial infected agents & $1$ & Number of agents in infectious state at the start of the simulation. \\
Chance of infection & $U(0,1)$ & Individual-level susceptibility parameter. \\
Chance of spreading & $U(0,1)$ & Individual-level contagiousness parameter. \\
Shedding rate & $U(0,1)$ & Controls magnitude of pathogen shedding over time. \\
\hline
\end{tabular}
\end{table}

\subsection{Simulation Parameters}

A set of behavioral and epidemiological parameters governs the agent-based simulation. The default framework values are summarized in Table~\ref{tab:sim-params}; the experimental values used in this study are reported in \reftable{tab:simulation_configuration}. Additional descriptions are provided below.

\paragraph{Defecation need parameters.}
Each agent is assigned a defecation rate drawn from a uniform distribution bounded by a lower and upper limit $(r_{\min}, r_{\max})$. This rate determines how quickly the agent transitions from \textit{no need to defecate} to \textit{urgent need to defecate}. After defecation, the fullness level is reset to its maximum and decreases over time based on the individualized rate. Once two internal thresholds are crossed, the agent moves sequentially through ``building pressure'', ``need to defecate'', and ``urgent need to defecate'' states. If defecation occurs while not in transport, pathogen shedding may take place within the current spatial unit. This design produces heterogeneous waste-generation behaviors that drive variability in pathogen deposition across the environment.

\paragraph{Disease progression parameters.}
Disease evolution follows a four-state progression: \textit{susceptible}, \textit{exposed}, \textit{infectious}, and \textit{recovered}. The durations of the exposed, infectious, and recovered states are determined by three parameters: incubation period $(D_E)$, infectious period $(D_I)$, and recovery period $(D_R)$. Transition is time-dependent, and agents return to the susceptible state after the recovery period, modeling short-term immunity and allowing reinfection.

\paragraph{Infection transmission parameters.}
Transmission is probabilistic and controlled by three main parameters: the infection ratio $(p_{\text{inf}})$, the number of secondary infections that can be attempted by an infectious agent per step $(n_{\text{spread}})$, and the number of infection attempts associated with a shared spatial unit $(n_{\text{unit}})$. During co-location events, the model evaluates a subset of nearby agents as potential transmission targets, enabling direct human-to-human spread across shared spatial units.

\paragraph{Pathogen shedding parameters.}
Infectious agents shed pathogens at a rate sampled from a uniform distribution in $[0, 1]$. The pathogen load is updated using a gamma-like shedding curve that increases early in the infectious period, peaks, and then declines during recovery. This supports temporal variation in shedding kinetics and captures diverse shedding profiles between individuals.

\subsection{Infectious Disease and Shedding Modeling}
\label{sec:infectious-disease}

To integrate disease dynamics within the patterns-of-life simulation, we developed an agent-based infectious disease model that captures the key mechanisms of disease transmission and infection progression at the individual-agent level. The model prioritizes both efficiency and realism, allowing disease spread conditional on daily agent interactions rather than being dictated by predefined mathematical equations (e.g., ordinary differential equations in an equation-based compartmental model). For each agent, the model tracks its disease status, which evolves over time according to contact events, probabilistic rules, and individual biological variability.

\subsubsection{Agent-based SEIR Framework}
We adapt the susceptible-exposed-infectious-recovered (SEIR) framework at the individual-agent level, with each agent transitioning among four states: Susceptible (S), Exposed (E), Infectious (I), and Recovered (R):
\[
S \rightarrow E \rightarrow I \rightarrow R \rightarrow S.
\]
Such a design enables simulation of disease dynamics at an individual level, allowing for heterogeneity in infection progression and shedding over time within the simulated population.
We consider a disease that is transmitted through the respiratory route while infected individuals also shed pathogen biomarkers in feces. Susceptible agents may become exposed after contact with infectious agents during social interactions. After exposure, they enter a latent phase, the ``Exposed'' state, in which they neither transmit the pathogen nor shed it in feces. After an individualized period, they transition to the Infectious state, during which they can transmit the disease and shed pathogen biomarkers. Finally, they recover and gain temporary immunity before potentially returning to the Susceptible state. This cyclic structure supports long-term simulations, although the long recovery period used in our experiments prevents reinfection within the simulated year.

\subsubsection{State Progression and Timing.}
The transmission of disease in the simulation follows a realistic, probabilistic process driven by direct agent interactions. 
At the start of the simulation, a predefined number of agents are initialized as infectious to seed the outbreak, while all others begin in the susceptible state. Agents in the \textit{Susceptible} state behave normally and do not contribute to pathogen shedding. As the simulation progresses, when a susceptible agent encounters an infectious one, the simulation model evaluates whether transmission can occur based on a series of probabilistic checks that govern infection likelihood and spread capacity.
To prevent uncontrolled outbreaks, each infectious agent is assigned a predefined infection limit that caps the number of individuals it can infect during its infectious period. If the agent has remaining capacity to transmit the disease, a random value is drawn from a uniform distribution and compared with the agent-specific transmission probability. If this initial draw succeeds, the susceptible target then undergoes a second probability check based on its agent-specific susceptibility.
These probabilistic checks capture individual-level variability in both transmitting and contracting infection, reflecting real-world differences in immune response, behavior, and lifestyle. Finally, a global disease-specific infection rate is applied to determine whether transmission occurs. Infection takes place only when all these probabilistic conditions are satisfied, ensuring that disease spread remains both stochastic and biologically realistic.

Once an agent becomes exposed, it remains in the \textit{Exposed} state for an individualized duration before transitioning to the \textit{Infectious} state. During the exposure period, the agent continues to move and interact with others but cannot yet transmit the disease. Each agent’s exposure duration is drawn from a global exposure threshold modulated by a smoothing factor that introduces slight variability across agents. After the exposure period ends, the agent becomes infectious and begins shedding pathogens into the system. This shedding activity contributes to the overall pathogen load in the simulated environment.
After remaining infectious for its assigned period, the agent transitions to the \textit{Recovered} state, during which it temporarily gains immunity. Once the recovery period is completed, the agent returns to the \textit{Susceptible} state and can once again participate in future infection cycles. This cyclical process of infection, recovery, and susceptibility enables the simulation to capture realistic epidemic waves and long-term population dynamics.

\subsubsection{Infection Layers of Interaction}
Infection in the simulation can occur through three interconnected layers of interaction that jointly capture both spatial and social aspects of disease spread. 

First, agents may infect others who are simply co-located in the same place, representing incidental contact driven by shared physical proximity. Second, agents who are co-located and actively attempting to expand their social networks can spread the infection to newly encountered individuals, capturing social-mixing behavior beyond familiar groups. Third, agents who are co-located, socially active, and interacting with close contacts such as friends, family members, or roommates can also transmit the disease while strengthening existing relationships. 

Together, these three layers, (1) co-location transmission, (2) contact-based transmission with social expansion, and (3) contact-based transmission with network strengthening, form a multi-level framework for simulating infection spread that reflects the intertwined spatial and social dimensions of real-world disease transmission.

\paragraph{(1) Co-location Transmission.}
One mode of disease spread in the simulation is based on co-location. When an agent detects that one of its basic needs is unmet, such as the need for social interaction (love and belonging), food, shelter, or defecation, it becomes active and initiates movement to satisfy that need. During this process, a co-location transmission mechanism is activated. A predefined parameter specifies the maximum number of agents that can be infected by a single infectious agent within the same location, preventing uncontrolled spread. This mechanism ensures that disease transmission occurs primarily among agents who are physically co-located and actively interacting, rather than indiscriminately affecting all agents in the vicinity. As agents move to fulfill their needs, they may encounter and infect others nearby, generating realistic patterns of spatially localized transmission that reflect everyday human behavior and movement dynamics.

\paragraph{(2) Contact-Based Transmission with Social Expansion.}
Another mode of transmission focuses on social relationships between agents, modeling infection spread through interactions within social networks. This mechanism captures infection events that occur when agents engage in social activities aimed at expanding their social connections. Such interactions represent close-contact scenarios, including social gatherings, dining with friends at restaurants, or meeting new individuals through shared activities. 

When multiple agents occupy the same location, an infectious agent does not attempt to infect everyone nearby; only those within close proximity are considered potential transmission targets. This approach produces a more realistic spread pattern by emphasizing that infection requires physical closeness; simply being in the same building or seeing another agent across the street does not lead to transmission. Additionally, the simulation incorporates randomness in social behavior, allowing agents to occasionally encounter and infect strangers outside their immediate social networks. This feature enables the pathogen to move beyond tightly clustered communities, capturing cross-group interactions and more accurately reflecting real-world patterns of disease transmission.

\paragraph{(3) Contact-Based Transmission with Network Strengthening.}
The third mode of transmission occurs when agents interact within their established social networks, reinforcing existing relationships with close contacts such as friends, family members, or roommates. In these interactions, infection can spread through repeated and prolonged contact, which reflects the higher transmission risk associated with close and frequent interactions in real-world settings. This layer represents stable, high-trust relationships where physical proximity and duration of contact are both elevated, making transmission more probable. By incorporating this mechanism, the simulation captures persistent infection chains within households and tight social circles, complementing the transient dynamics of social expansion and the incidental nature of co-location transmission.

\subsubsection{Mathematical Representation}
At the core of the infection mechanism is a probabilistic model that governs both transmission and state progression. When a susceptible agent encounters an infectious one, infection occurs only if multiple independent probability checks succeed:
\[
u_1 < p_{\text{spread}}, \quad
u_2 < p_{\text{infect}}, \quad
u_3 < \rho, \quad
\text{and } c_i < N_{\max},
\]
where $u_1, u_2, u_3 \sim U(0,1)$ are random values drawn from a uniform distribution, $p_{\text{spread}}$ is the infectious agent’s ability to transmit, $p_{\text{infect}}$ is the susceptibility of the recipient, $\rho$ is the global infection ratio, $c_i$ is the source agent’s current infection count, and $N_{\max}$ is the maximum number of infections allowed per source agent. Only when all four conditions hold does the susceptible agent become exposed.

Suppose that agent $i$ can infect at most $N_{\max}$ other agents during its infectious period. For each meeting between an infectious source $i$ and a susceptible recipient $r$, infection occurs according to a factored probability model:
\[
\Pr(r\ \text{becomes } E \mid i\ \text{meets } r) 
= \mathbf{1}\!\{\text{spread count} < N_{\max}\} \times 
p_{\text{spread}} \times p_{\text{infect}} \times \rho.
\]
Here, $p_{\text{spread}}$ represents the infectious agent’s transmissibility, $p_{\text{infect}}$ the susceptibility of the recipient, and $\rho$ a global infection ratio that regulates the baseline transmissibility of the disease. Each contact event is independent, and a successful transmission moves $r$ to the exposed state while recording $i$ as the infection source.

Once infected, each agent $a$ maintains individualized durations for the exposed, infectious, and recovered states: $\theta_E(a)$, $\theta_I(a)$, and $\theta_R(a)$. These values determine how long the agent remains in each state before transitioning. Let $d_E(a)$, $d_I(a)$, and $d_R(a)$ represent the elapsed time in each state. The state transitions follow:
\[
E \rightarrow I \text{ if } d_E(a) \geq \theta_E(a), \quad
I \rightarrow R \text{ if } d_I(a) \geq \theta_I(a), \quad
R \rightarrow S \text{ if } d_R(a) \geq \theta_R(a).
\]
When an agent reverts to $S$, all infection-related counters and identifiers are reset, ensuring that subsequent infections are treated as new and independent events.

\paragraph{Introducing Heterogeneity.}
To avoid synchronized state transitions and to better reflect real-world variability, we introduce heterogeneity through a smoothness parameter $s \in [0,1]$ drawn from a uniform distribution $s \sim U(0,1)$. This parameter perturbs the base durations $\bar{\theta}_E$, $\bar{\theta}_I$, and $\bar{\theta}_R$, which represent the global durations for the exposed, infectious, and recovered states. The individualized durations for each agent are computed as:
\[
\tilde{\theta} = \left\lceil \bar{\theta} + (0.5 - s)\bar{\theta} \right\rceil.
\]
Here, $\bar{\theta}$ denotes the global base value and $\tilde{\theta}$ the agent-specific adjusted value. The randomization ensures that agents transition at different times, producing smoother epidemic curves and more realistic infection dynamics across the simulated population.

\paragraph{Pathogen Shedding Dynamics.}
Each infected agent begins shedding pathogen biomarkers after certain period of time after infection, which was assumed to be the period that agent is infectious in this simulation. The pathogen load over time is modeled using a gamma-like function that captures a rise-and-decay pattern. The number of pathogen units shed at $t$ days after the agent becomes infectious is:
\begin{equation}
N(t) = N_0\, t^{b} e^{-a t}.
\end{equation}
where $N_0$ is the shedding scale at the start of the infectious period, $a$ is the decay-rate parameter, and $b$ is the shape parameter. We use $a=2$, $b=8$, and $N_0=10^7$ in the simulation.

\paragraph{Simulation Update Cycle.}
The infection model operates in a discrete-time manner, where each simulation tick represents one unit of simulated time (e.g., 5 minutes). During each tick, every agent independently updates its internal disease state and interacts with other agents, synchronizing biological processes with social behavior. Specifically, at each step, the agent increments the time spent in its current health state ($d_E$, $d_I$, or $d_R$), updates its pathogen load if it is in the infectious state, and evaluates transition conditions based on the corresponding thresholds $\theta_E$, $\theta_I$, and $\theta_R$. This iterative process allows infection dynamics to evolve continuously and organically as agents move, meet, and change states throughout the simulation.

\section{Results}
\label{sec:results}

In this section, we present the environmental setup used for the simulation experiments along with the resulting outcomes of the simulation.
\begin{table}[t]
    \caption{Key configuration parameters used in the simulation experiments}
    \label{tab:simulation_configuration}
    \centering
    \begin{tabular}{|l | l | p{6cm}|}
        \hline
        \textbf{Parameter}                      & \textbf{Value}            & \textbf{Description} \\ \hline
        \texttt{numOfAgents}                    & 10,000                    & Total number of simulated agents. \\ \hline
        \texttt{numberOfDaysToBeExposed}        & 7                         & Exposure period in days.\\ \hline
        \texttt{numberOfDaysToBeInfectious}     & 14                        & Infectious period in days. \\ \hline
        \texttt{numberOfDaysToBeRecovered}      & 2800                      & Recovery duration in days. \\ \hline
        \texttt{numberOfSpreadPerPerson}        & 50                        & Maximum number of infection attempts per infectious individual. \\ \hline
        \texttt{numberOfSpreadPerUnit}          & 10                        & Maximum number of infection attempts per building unit per agent. \\ \hline
        \texttt{numberOfInitialInfectedAgents}  & 10                        & Number of infected agents at simulation start. \\ \hline
        \texttt{infectionRatio}                 & 0.1--0.5                & Infection rates evaluated in increments of 0.025. \\ \hline
    \end{tabular}
\end{table}

\subsection{Environmental Setup and Simulation Configuration}

All simulations were executed on a system equipped with an 11th Gen Intel(R) Core(TM) i5-1135G7 CPU @ 2.40~GHz running Fedora Linux 42 (Workstation Edition) with 16~GiB of RAM.  
The simulation environment is defined by more than 60 parameters; we present the most relevant ones in \reftable{tab:simulation_configuration}.  
For this study, we conducted simulations with 10,000 agents within Fulton County, Georgia, USA. This population size was selected to represent an upstream catchment-scale setting rather than the full population served by a large WWTP. The experiment demonstrates the framework’s ability to resolve individual mobility, defecation events, toilet-use locations, and pathogen shedding at fine spatial and temporal scales. Each simulation covered approximately 348 days (100,000 five-minute ticks) and required about 10 hours to complete. Larger populations can be represented by increasing the number of agents, although doing so increases computational cost.

\subsubsection*{Scalability and computational considerations}
The Fulton County experiment uses 10,000 agents as a reproducible case study rather than as the maximum supported population size of the framework. This setting was selected because the experiment simulates approximately one year at five-minute resolution and is repeated across multiple infection-rate scenarios. In our experiments, each 10,000-agent scenario covered about 348 simulation days and required approximately 10 hours on a laptop-class machine with 16 GB of memory. Thus, the current configuration is intended to demonstrate the integrated wastewater extension under a manageable computational budget, while preserving individual-level mobility, social interaction, defecation, infection, and shedding dynamics. Although the agents follow behaviorally realistic mobility patterns, downsampling Fulton County’s population of approximately 1,070,000 people to 10,000 agents may affect the simulated disease dynamics. Our goal is not to reproduce a specific epidemic exactly, but to demonstrate how mobility and toilet use affect wastewater-based epidemiology; the 10,000-agent case study is sufficient for that purpose.

%The scalability of the framework comes from the underlying HD-GEN/POL implementation~\cite{amiri2026hd}. The enhanced POL engine has been evaluated at larger population sizes and was able to complete simulations with more than 150,000 agents, while the original implementation did not scale beyond approximately 15,000 agents in the reported experiments. These improvements were achieved through optimized initialization, reduced redundant logging, faster nearest-location calculations, and support for parallel execution of independent simulation runs. For operational city-scale deployments, the framework can be scaled in three ways: (1) increasing the number of simulated agents directly, (2) using a weighted synthetic population where each agent represents multiple residents, and (3) decomposing the study region into neighborhoods or sewersheds that can be simulated and analyzed in parallel. Full one-to-one simulation of an entire metropolitan population, combined with hydraulic sewer transport and calibration to empirical wastewater data, remains an important direction for future work.

\subsection{Experimental Results}

We evaluated the model using infection rates from 0.1 to 0.5 with increments of 0.025. The complete collection of results, including videos and plots, is available on GitHub. This paper presents four representative plots that demonstrate the key findings, while the full artifacts can be accessed online.

\subsubsection{Disease Dynamics}
\begin{figure}[htbp]
    \centering
    \begin{subfigure}[b]{0.49\linewidth}
        \centering
        \includegraphics[width=\linewidth]{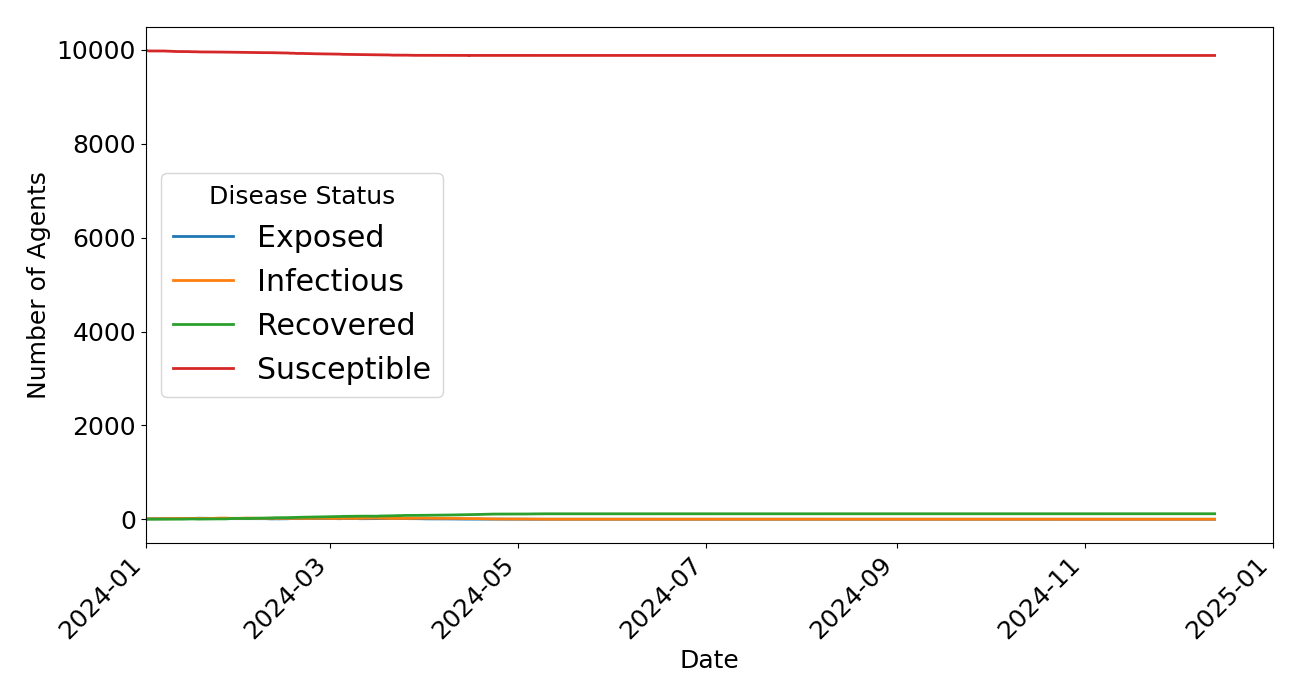}
        \caption{Lowest transmission scenario (Infection rate = 0.1)}
        \label{subfig:fulton_ir_0_1}
    \end{subfigure}
    \begin{subfigure}[b]{0.49\linewidth}
        \centering
        \includegraphics[width=\linewidth]{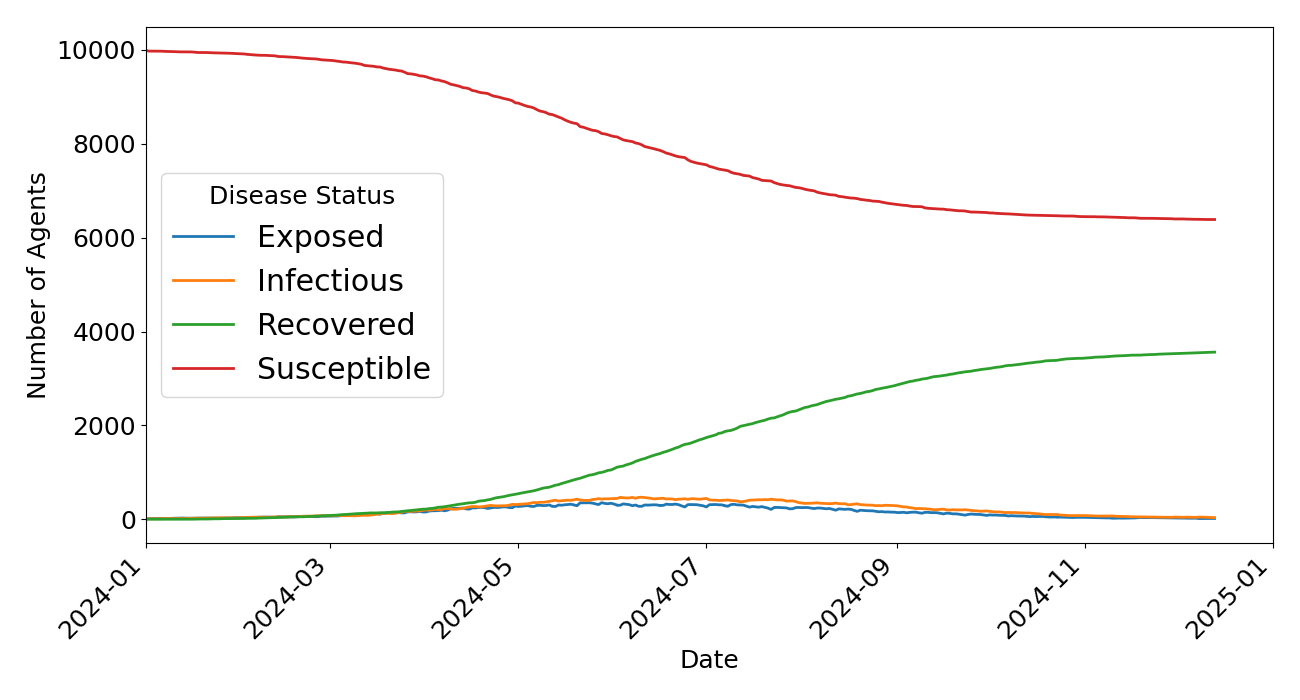}
        \caption{Slightly higher transmission scenario (Infection rate = 0.15)}
        \label{subfig:fulton_ir_0_15}
    \end{subfigure}

    \begin{subfigure}[b]{0.49\linewidth}
        \centering
        \includegraphics[width=\linewidth]{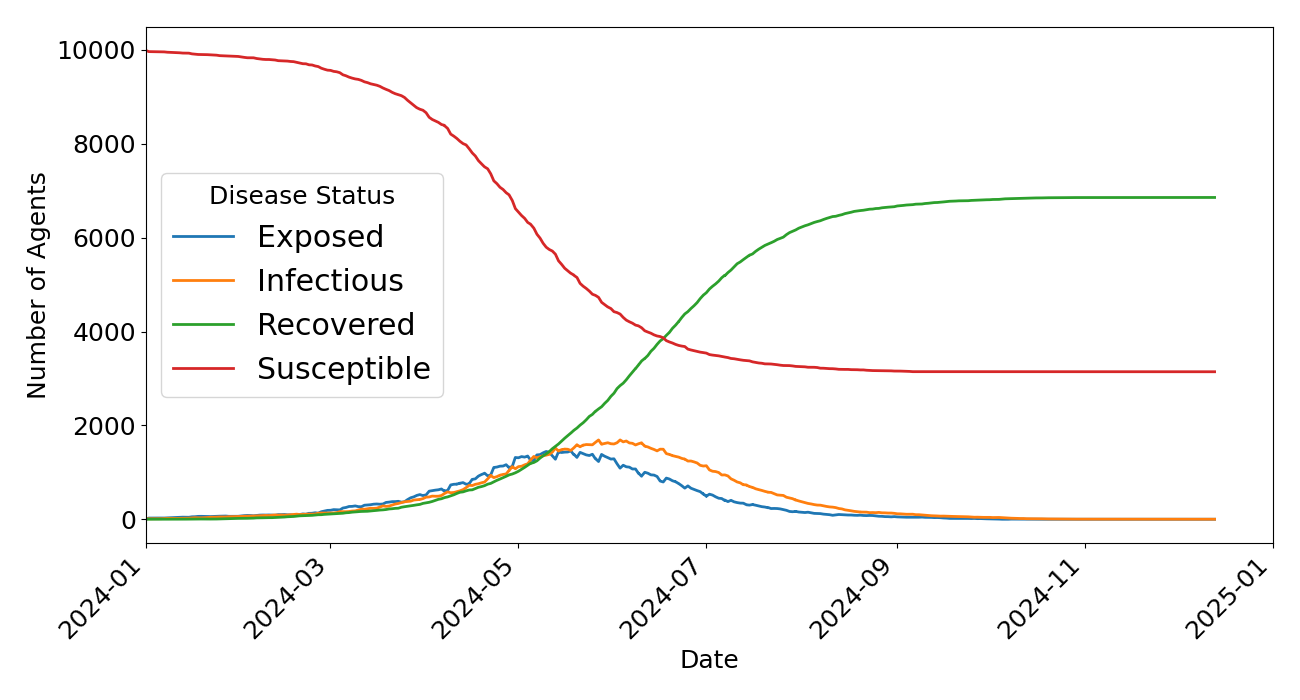}
        \caption{Moderate transmission scenario (Infection rate = 0.25)}
        \label{subfig:fulton_ir_0_25}
    \end{subfigure}
    \begin{subfigure}[b]{0.49\linewidth}
        \centering
        \includegraphics[width=\linewidth]{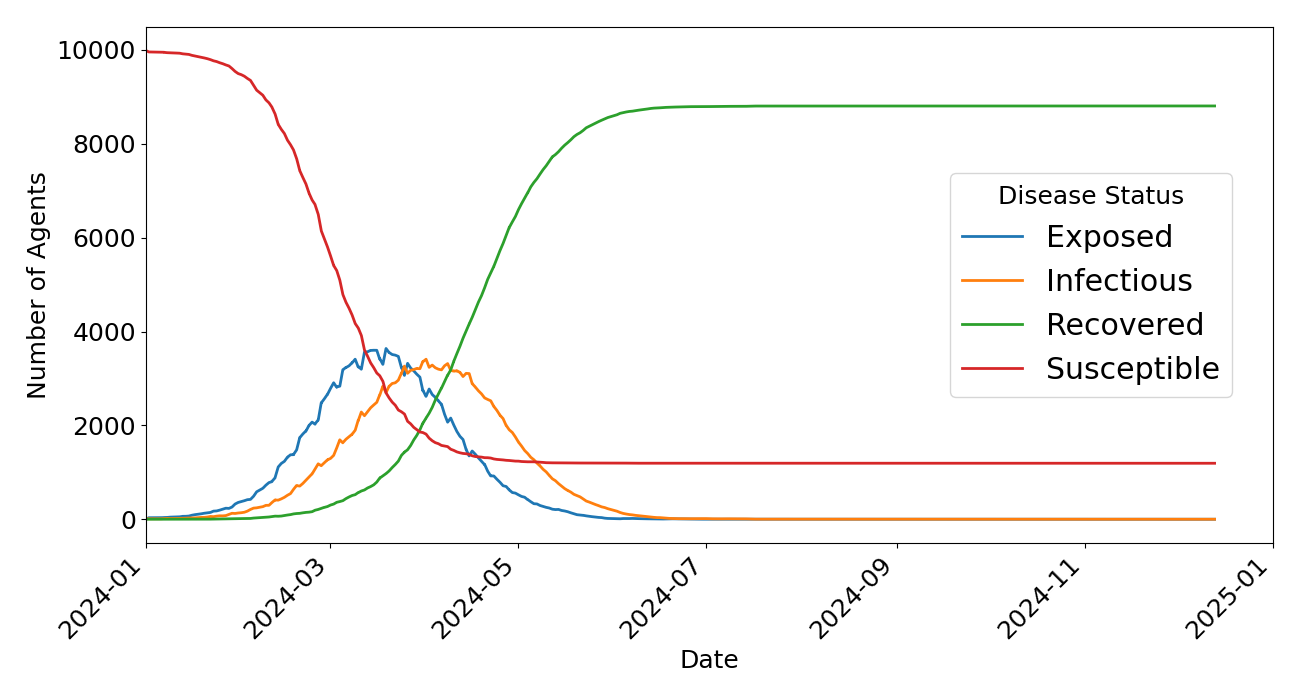}
        \caption{Highest transmission scenario (Infection rate = 0.5)}
        \label{subfig:fulton_ir_0_5}
    \end{subfigure}
  \vspace{-0.3cm}
    \caption{Disease progression over time in the Fulton County 10K simulation across different infection rates. \fullresultsnote \vspace{-0.3cm}}\vspace{-0.3cm}
    \Description{Four line charts showing susceptible, exposed, infectious, and recovered agent counts over time for infection rates 0.1, 0.15, 0.25, and 0.5.}
    \label{fig:disease_status_fulton_infection_rates}
\end{figure}

\reffig{fig:disease_status_fulton_infection_rates} illustrates the temporal dynamics of disease progression for simulations conducted with varying infection rates (0.1, 0.15, 0.25, and 0.5) in a population of 10,000 agents in Fulton County. Each subplot depicts the transitions among key disease states—susceptible, exposed, infectious, and recovered. As infection rate increases, the spread accelerates, leading to higher and earlier peaks in the infectious population and a faster decline in susceptible individuals. This pattern highlights how elevated transmission rates intensify outbreak severity and shorten epidemic duration. 
Notably, we assume a long recovery period so individuals do not become susceptible again immediately. This prevents disease spread during recovery in the SEIR model, instead of allowing reinfection after only a short recovery.
\reffig{subfig:fulton_ir_0_1}:
In this scenario, only a small portion of the population transitions from susceptible to infected. The curves for exposed and infectious agents remain low throughout the simulation period, indicating limited disease spread. The susceptible population shows only a minimal decrease over time. As a result, recovered individuals constitute a very small fraction of the population by the end of the year.
\reffig{subfig:fulton_ir_0_15}:
With an increased transmission rate, the epidemic expands more noticeably and reaches a higher peak of infectious individuals earlier in the year. The susceptible population gradually decreases as more individuals become exposed and later infectious. The recovered population grows steadily as immunity accumulates. Although the outbreak is more prominent than in the previous case, a substantial majority of agents remain susceptible by the end of the simulation.
\reffig{subfig:fulton_ir_0_25}:
At this higher infection rate, the disease spreads rapidly through the population. The infectious and exposed curves rise sharply, reflecting a stronger and faster outbreak. The susceptible pool diminishes significantly as recovery rates increase and immunity becomes widespread. By the end of the year, the epidemic has affected a large portion of the population, leading to a dominant recovered group.
\reffig{subfig:fulton_ir_0_5}:
Under aggressive transmission conditions, the epidemic unfolds very quickly. The susceptible population plummets early, as the infection spreads nearly unchecked through the community. Peaks in the exposed and infectious populations occur rapidly but are short-lived due to the rapid exhaustion of susceptible agents. Eventually, the recovered population stabilizes at a very high level, indicating near-complete epidemic saturation.

\subsubsection{Epidemic Curve Analysis}
\begin{figure}[htbp]
    \centering

    \begin{subfigure}[b]{0.49\linewidth}
        \centering
        \includegraphics[width=\linewidth]{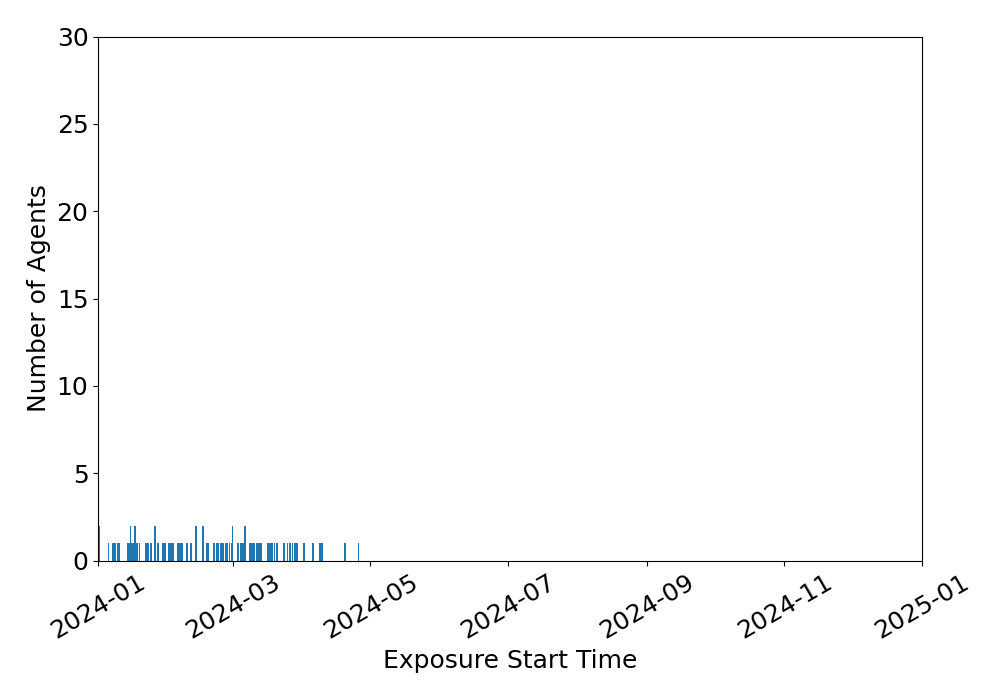}
        \caption{Lowest transmission scenario (Infection rate = 0.1)}
        \label{subfig:epicurve_ir_0_1}
    \end{subfigure}
    \begin{subfigure}[b]{0.49\linewidth}
        \centering
        \includegraphics[width=\linewidth]{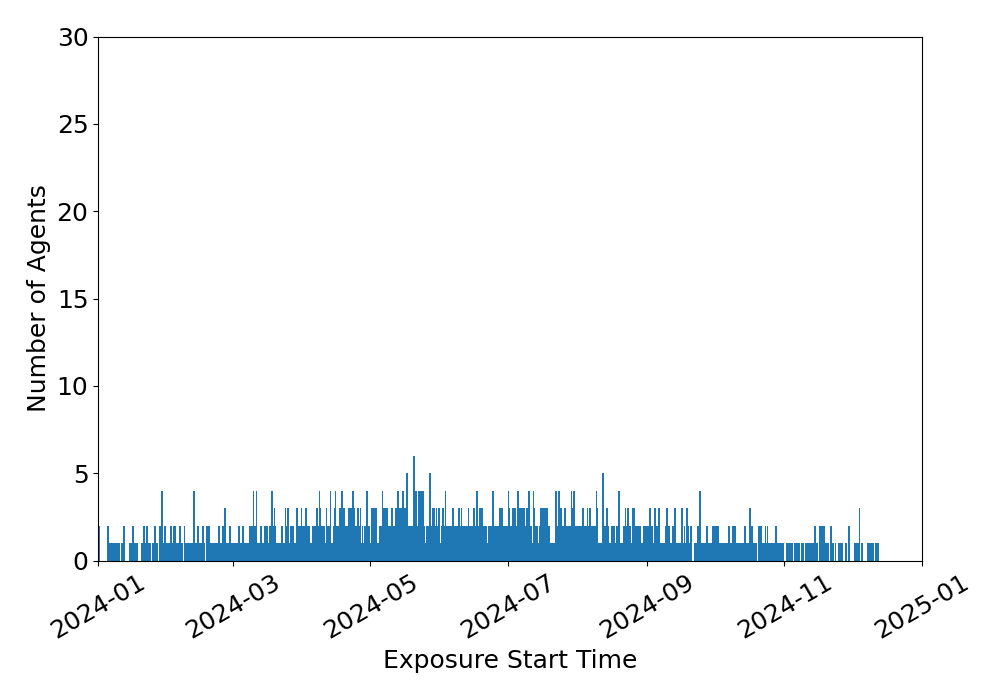}
        \caption{Slightly higher transmission scenario (Infection rate = 0.15)}
        \label{subfig:epicurve_ir_0_15}
    \end{subfigure}

    \begin{subfigure}[b]{0.49\linewidth}
        \centering
        \includegraphics[width=\linewidth]{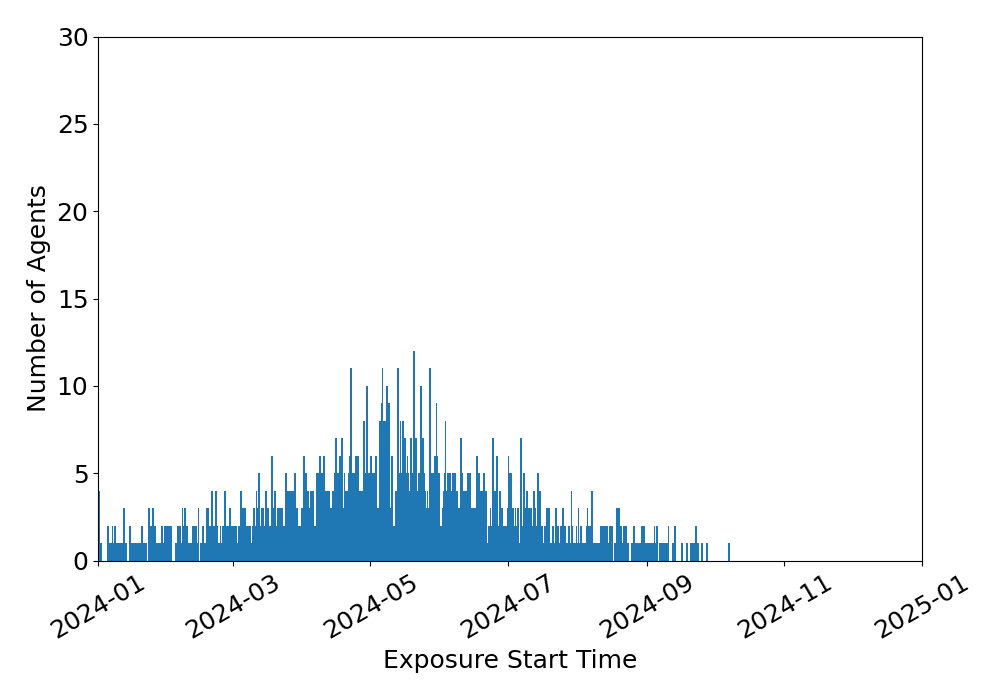}
        \caption{Moderate transmission scenario (Infection rate = 0.25)}
        \label{subfig:epicurve_ir_0_25}
    \end{subfigure}
    \begin{subfigure}[b]{0.49\linewidth}
        \centering
        \includegraphics[width=\linewidth]{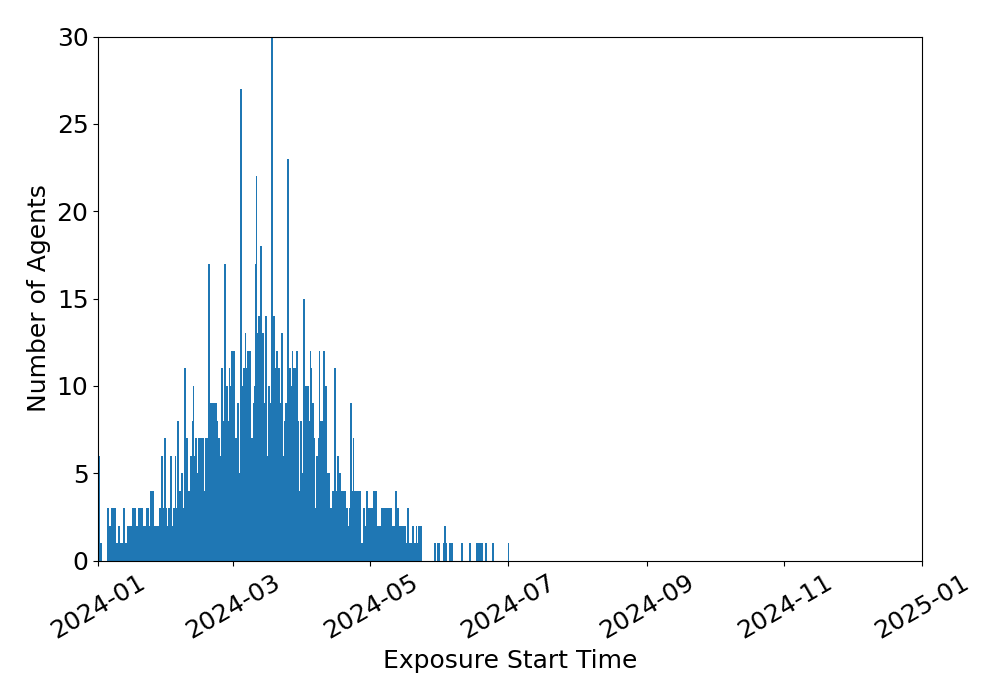}
        \caption{Highest transmission scenario (Infection rate = 0.5)}
        \label{subfig:epicurve_ir_0_5}
    \end{subfigure}
    \vspace{-0.3cm}
    \caption{Daily number of new cases in the Fulton County 10K simulation across different infection rates. \vspace{-0.3cm} }\vspace{-0.3cm}
    \Description{Four epidemic curves showing daily new cases over time for infection rates 0.1, 0.15, 0.25, and 0.5.}
    \label{fig:epidemic_curve_fulton_infection_rates}
\end{figure}

\reffig{fig:epidemic_curve_fulton_infection_rates} presents the epidemic curves showing the daily number of new cases for simulations conducted with varying infection rates (0.1, 0.15, 0.25, and 0.5) in a population of 10,000 agents in Fulton County. Each subplot captures the temporal pattern of the outbreak for a specific infection probability, highlighting changes in peak intensity, outbreak duration, and overall epidemic behavior. As infection rates rise, the system transitions from minimal spread to rapid, high-intensity epidemic waves. These contrasts demonstrate the critical role of transmission probability in determining whether an outbreak remains contained or expands rapidly through the population.
\reffig{subfig:epicurve_ir_0_1}:
With an infection rate of 0.1, new case numbers remain consistently low throughout the simulation period. Transmission fails to build momentum, resulting in only small clusters of infections that fade quickly. The absence of any pronounced peak reflects insufficient spread to drive sustained epidemic growth. Most individuals remain uninfected, indicating an effectively self-limiting outbreak.
\reffig{subfig:epicurve_ir_0_15}:
At a transmission rate of 0.15, the outbreak becomes more evident and persists longer. Daily new cases rise gradually to a modest peak, showing that infections can propagate but still at limited speed. Although the epidemic is more sustained than in the lowest scenario, the number of newly infected agents remains relatively small, and transmission eventually subsides without overwhelming the population.
\reffig{subfig:epicurve_ir_0_25}:
With an infection rate of 0.25, the epidemic progresses much more aggressively. The case curve rises sharply, reaching a distinct peak as infections spread efficiently through the population. Following this rapid growth, daily case counts decline as susceptible individuals are depleted. This behavior is characteristic of a classical epidemic wave driven by strong transmission and expanding immunity.
\reffig{subfig:epicurve_ir_0_5}:
At the infection rate of 0.5, the outbreak escalates very quickly. New case counts surge early in the year, producing a short yet severe epidemic wave. Once susceptibility is exhausted, transmission collapses abruptly, and the epidemic ends rapidly. This scenario shows how high transmission induces intense but short-lived outbreaks due to rapid depletion of susceptible agents.
These comparisons confirm that increases in infection rate amplify the speed and magnitude of epidemic spread, producing earlier and sharper peaks while shortening the overall outbreak duration.

\subsubsection{Temporal Variation in Pathogen Load}
\begin{figure}[htbp]
    \centering

    \begin{subfigure}[b]{0.49\linewidth}
        \centering
        \includegraphics[width=\linewidth]{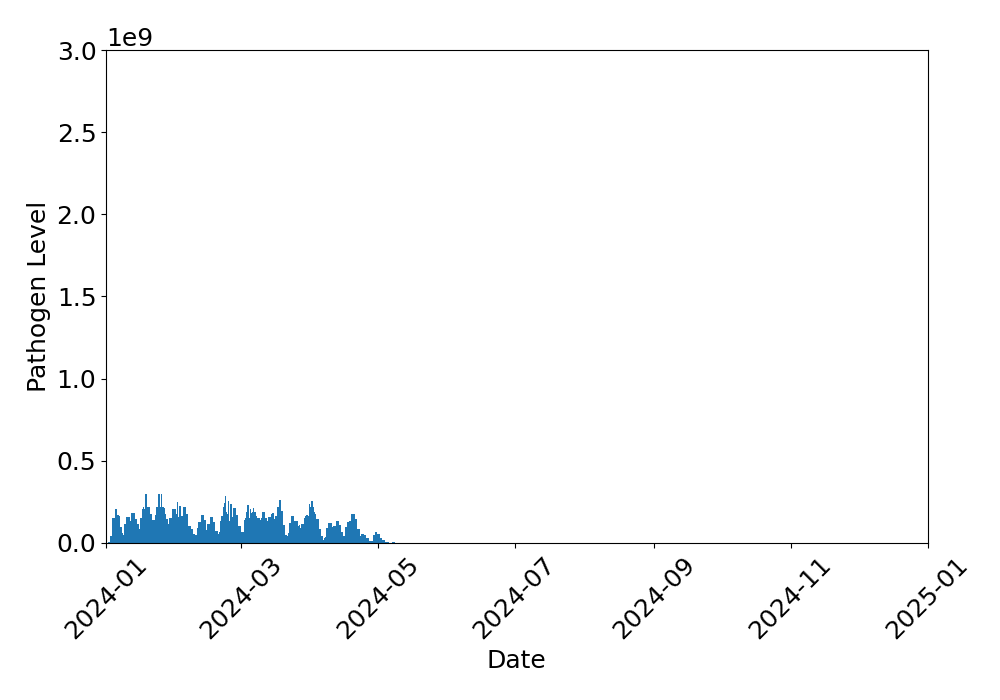}
        \caption{Lowest transmission scenario (Infection rate = 0.1)}
        \label{subfig:pathogen_ir_0_1}
    \end{subfigure}
    \begin{subfigure}[b]{0.49\linewidth}
        \centering
        \includegraphics[width=\linewidth]{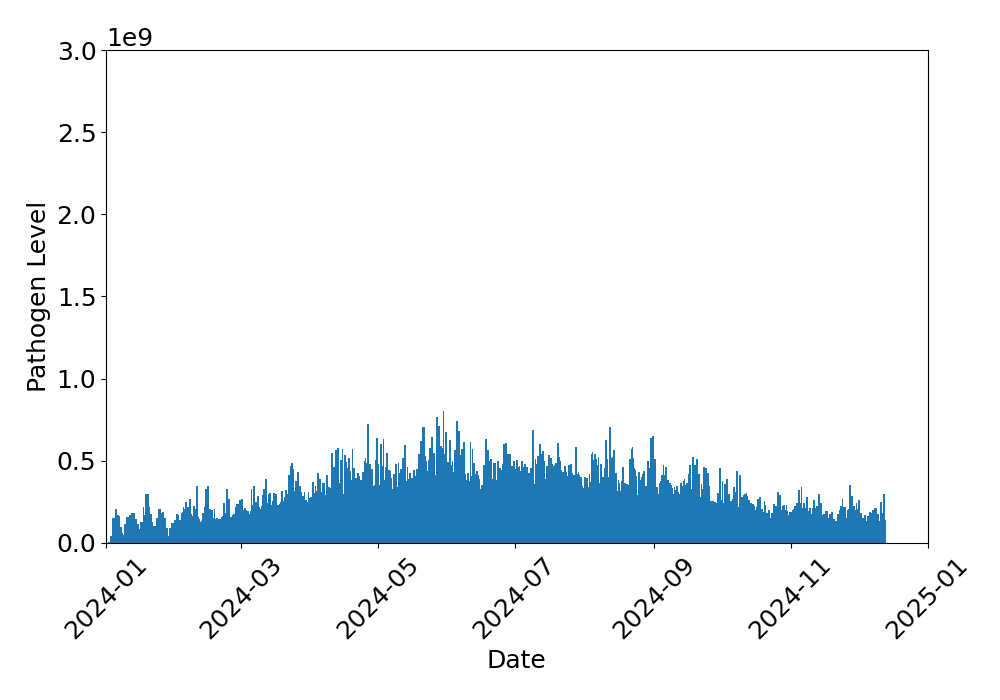}
        \caption{Slightly higher transmission scenario (Infection rate = 0.15)}
        \label{subfig:pathogen_ir_0_15}
    \end{subfigure}

    \begin{subfigure}[b]{0.49\linewidth}
        \centering
        \includegraphics[width=\linewidth]{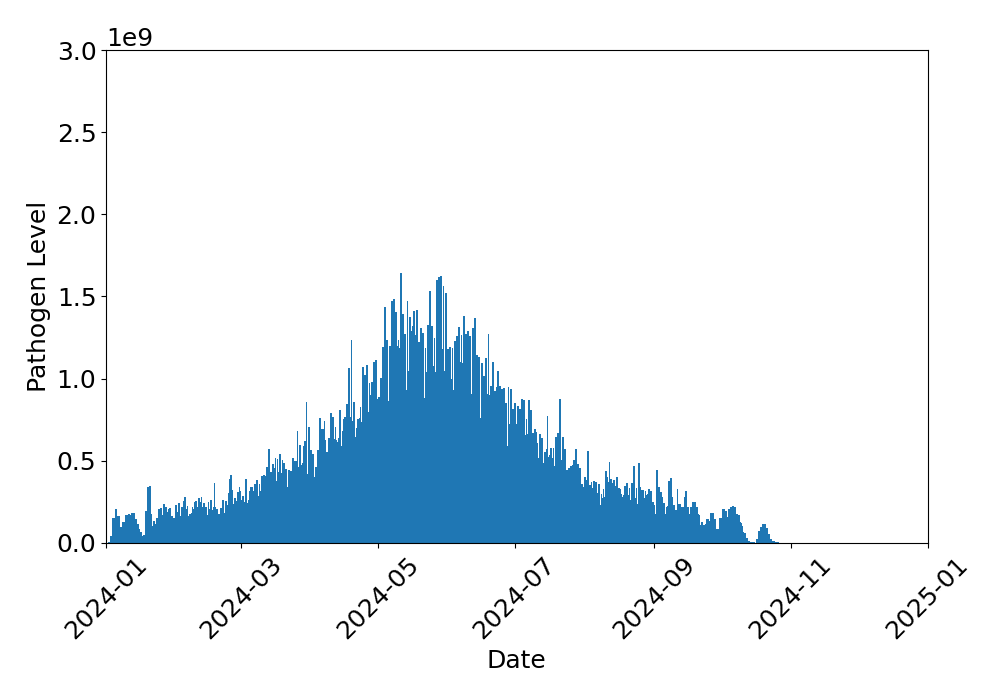}
        \caption{Moderate transmission scenario (Infection rate = 0.25)}
        \label{subfig:pathogen_ir_0_25}
    \end{subfigure}
    \begin{subfigure}[b]{0.49\linewidth}
        \centering
        \includegraphics[width=\linewidth]{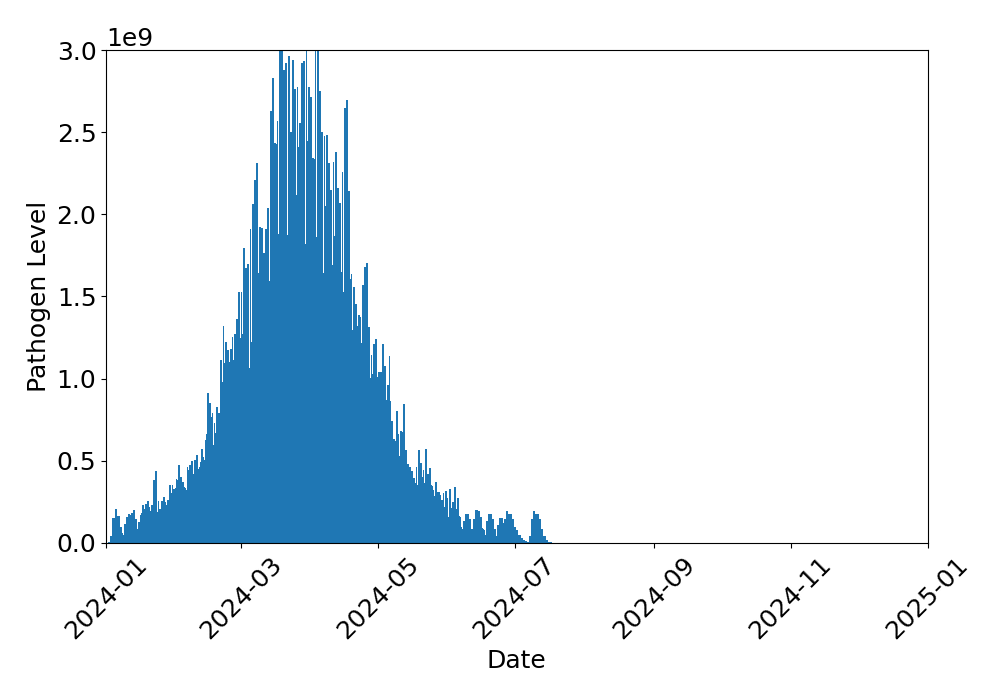}
        \caption{Highest transmission scenario (Infection rate = 0.5)}
        \label{subfig:pathogen_ir_0_5}
    \end{subfigure}
    \vspace{-0.3cm}
    \caption{Daily total pathogen load in wastewater for the Fulton County 10K simulation across different infection rates. \fullresultsnote \vspace{-0.3cm}}\vspace{-0.3cm}
    \Description{Four line charts showing total daily wastewater pathogen load over time for infection rates 0.1, 0.15, 0.25, and 0.5.}
    \label{fig:pathogen_load_fulton_infection_rates}
\end{figure}

\reffig{fig:pathogen_load_fulton_infection_rates} shows the total daily pathogen load in wastewater for simulations with varying infection rates (0.1, 0.15, 0.25, and 0.5) in a population of 10,000 agents in Fulton County. The temporal trends closely mirror the epidemic dynamics, since pathogen shedding depends on ongoing infections. As transmission increases, the wastewater pathogen load becomes stronger, peaks earlier, and declines faster as susceptible individuals are depleted. This relationship highlights how wastewater-based surveillance provides a nonintrusive measure of outbreak scale and timing.
\reffig{subfig:pathogen_ir_0_1}:
With an infection rate of 0.1, pathogen levels in wastewater remain very low and fluctuate only briefly during the initial months. The simulated signal disappears by late spring, indicating that the outbreak is too small and short-lived to sustain an elevated wastewater pathogen load. The data indicate minimal community spread and rapid failure of the pathogen to establish sustained transmission.
\reffig{subfig:pathogen_ir_0_15}:
At a transmission rate of 0.15, daily pathogen loads increase gradually as infections persist longer and occur in greater numbers. The peak remains moderate, but the presence of a clear curve extending into late summer reflects a more prolonged and noticeable outbreak. Although still controlled relative to higher transmission scenarios, the simulated wastewater signal persists for much of the year.
\reffig{subfig:pathogen_ir_0_25}:
With an infection rate of 0.25, the wastewater pathogen load increases sharply. Pathogen levels peak at high concentrations in late spring, matching the accelerated spread observed in disease dynamics. The pathogen load then declines steadily as herd immunity builds and infection prevalence falls. This strong environmental signature is characteristic of a substantial and rapidly expanding epidemic wave.
\reffig{subfig:pathogen_ir_0_5}:
In the most aggressive transmission setting, pathogen loads escalate extremely fast and reach the highest values recorded among all simulations. The peak occurs early, followed by a rapid collapse once the susceptible pool is nearly exhausted. The brief but intense spike indicates widespread transmission compressed into a short period, making wastewater levels a highly sensitive and timely indicator of severe outbreaks.
Overall, these results underscore that wastewater pathogen load serves as a robust indicator of epidemic scale, rising in tandem with infection prevalence and providing early warning of rapid outbreak growth.

\subsubsection{Spatial Distribution of Pathogen Spread}

\begin{figure}[!ht]
    \centering
    \begin{subfigure}[b]{0.49\linewidth}
        \centering
        \includegraphics[width=\linewidth]{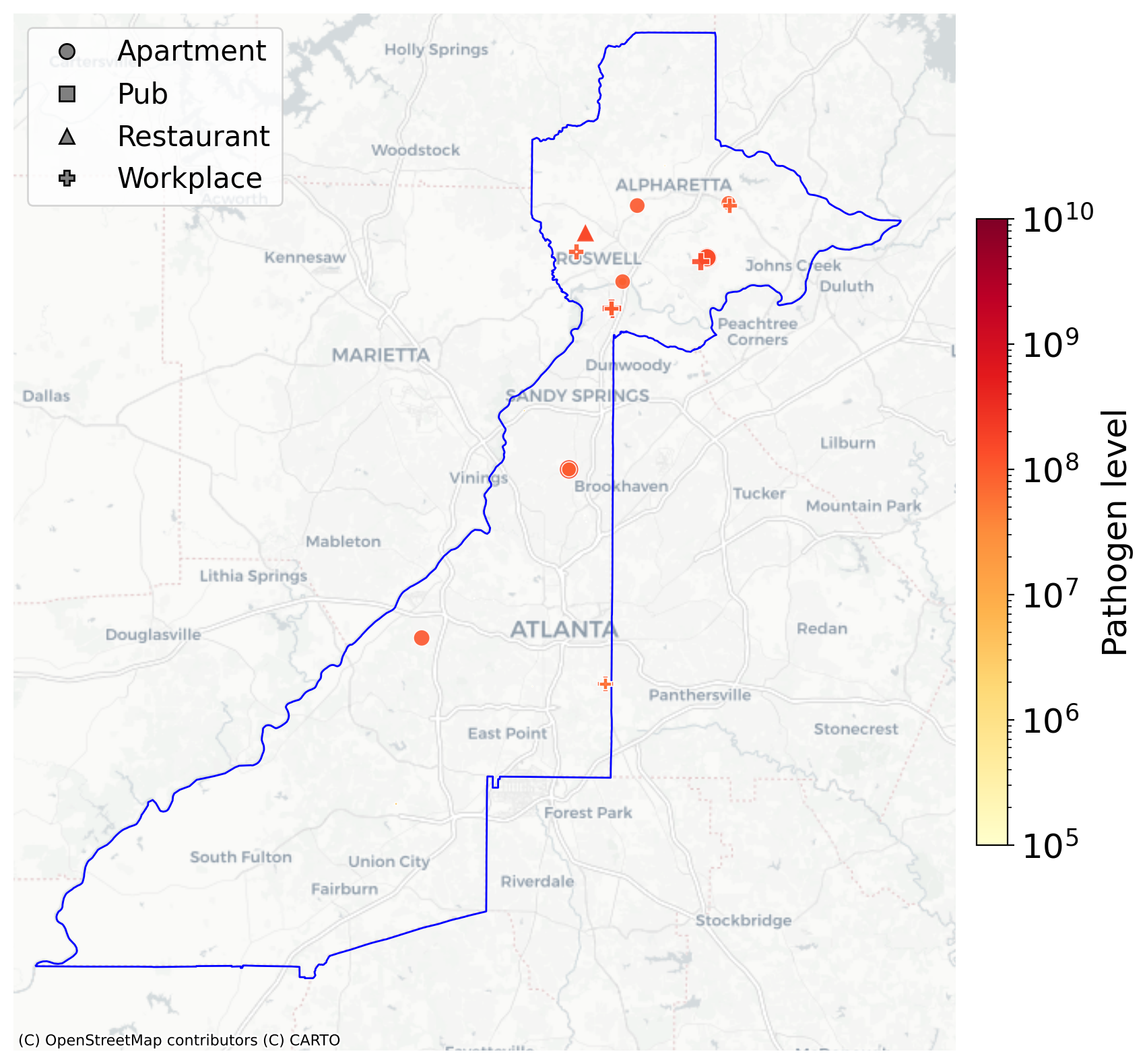}
        \caption{Lowest transmission scenario (Infection rate = 0.1)}
        \label{subfig:spatial_feb1_ir_0_1}
    \end{subfigure}
    \begin{subfigure}[b]{0.49\linewidth}
        \centering
        \includegraphics[width=\linewidth]{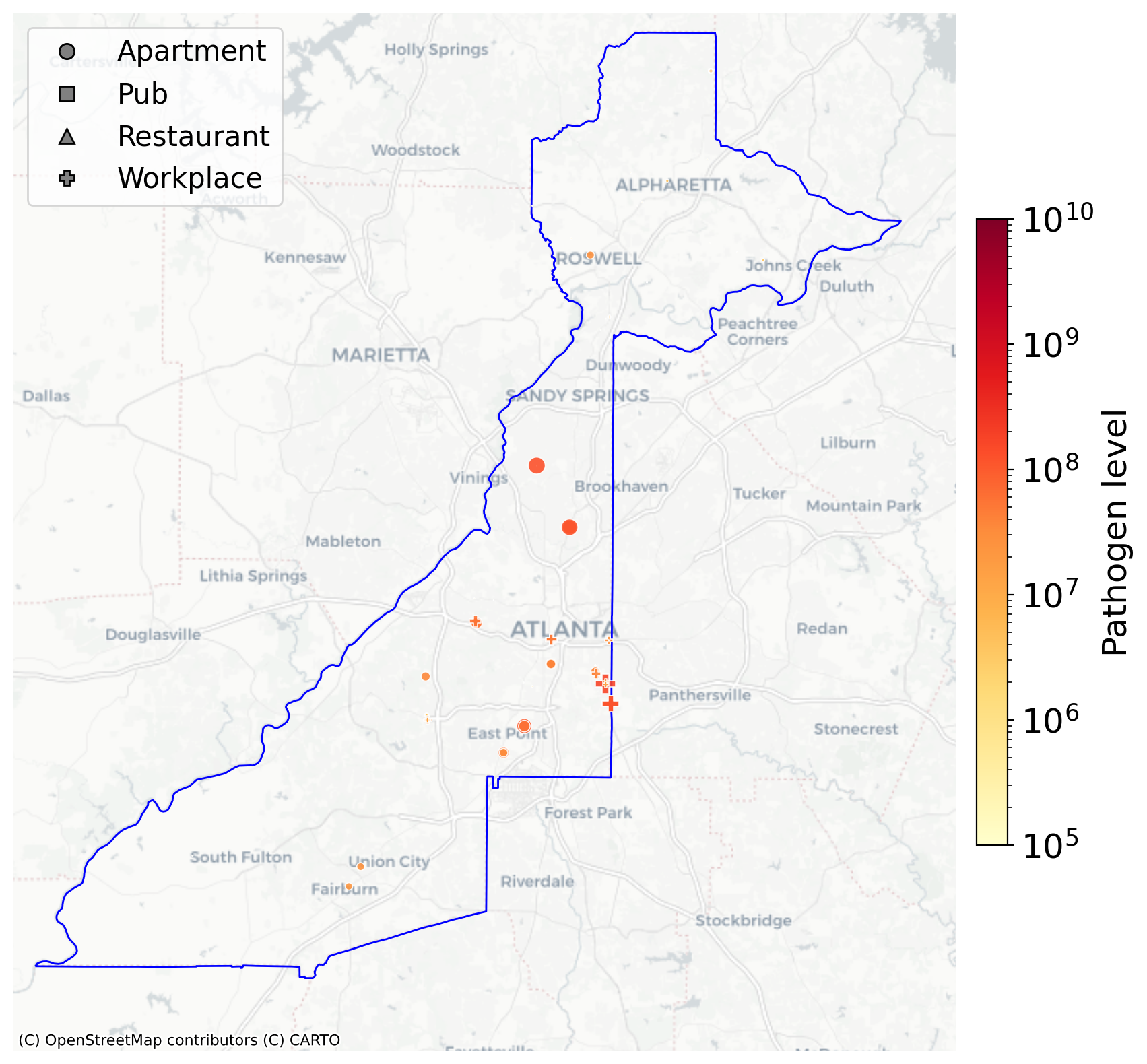}
        \caption{Slightly higher transmission scenario (Infection rate = 0.15)}
        \label{subfig:spatial_feb1_ir_0_15}
    \end{subfigure}
    \begin{subfigure}[b]{0.49\linewidth}
        \centering
        \includegraphics[width=\linewidth]{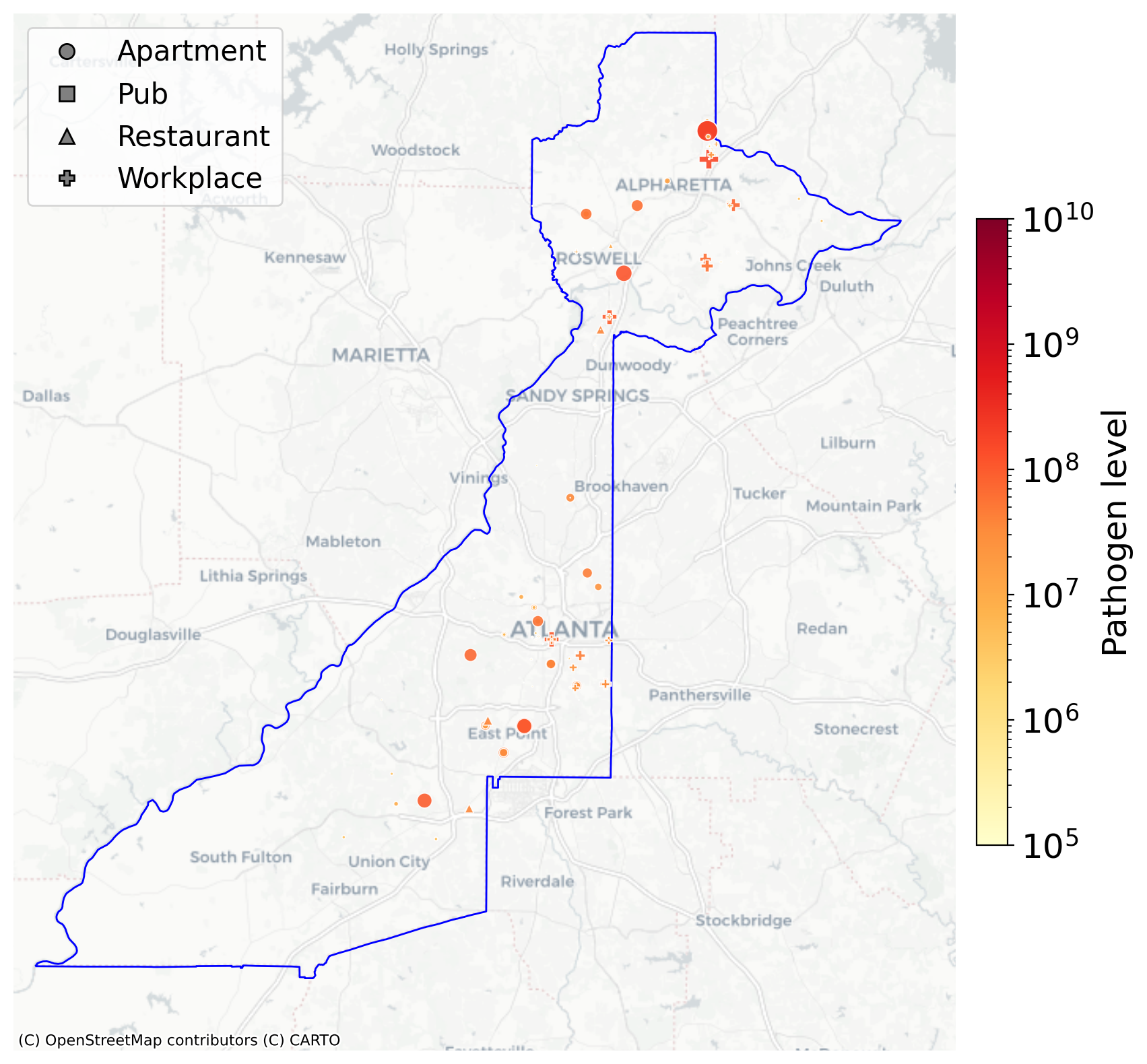}
        \caption{Moderate transmission scenario (Infection rate = 0.25)}
        \label{subfig:spatial_feb1_ir_0_25}
    \end{subfigure}
    \begin{subfigure}[b]{0.49\linewidth}
        \centering
        \includegraphics[width=\linewidth]{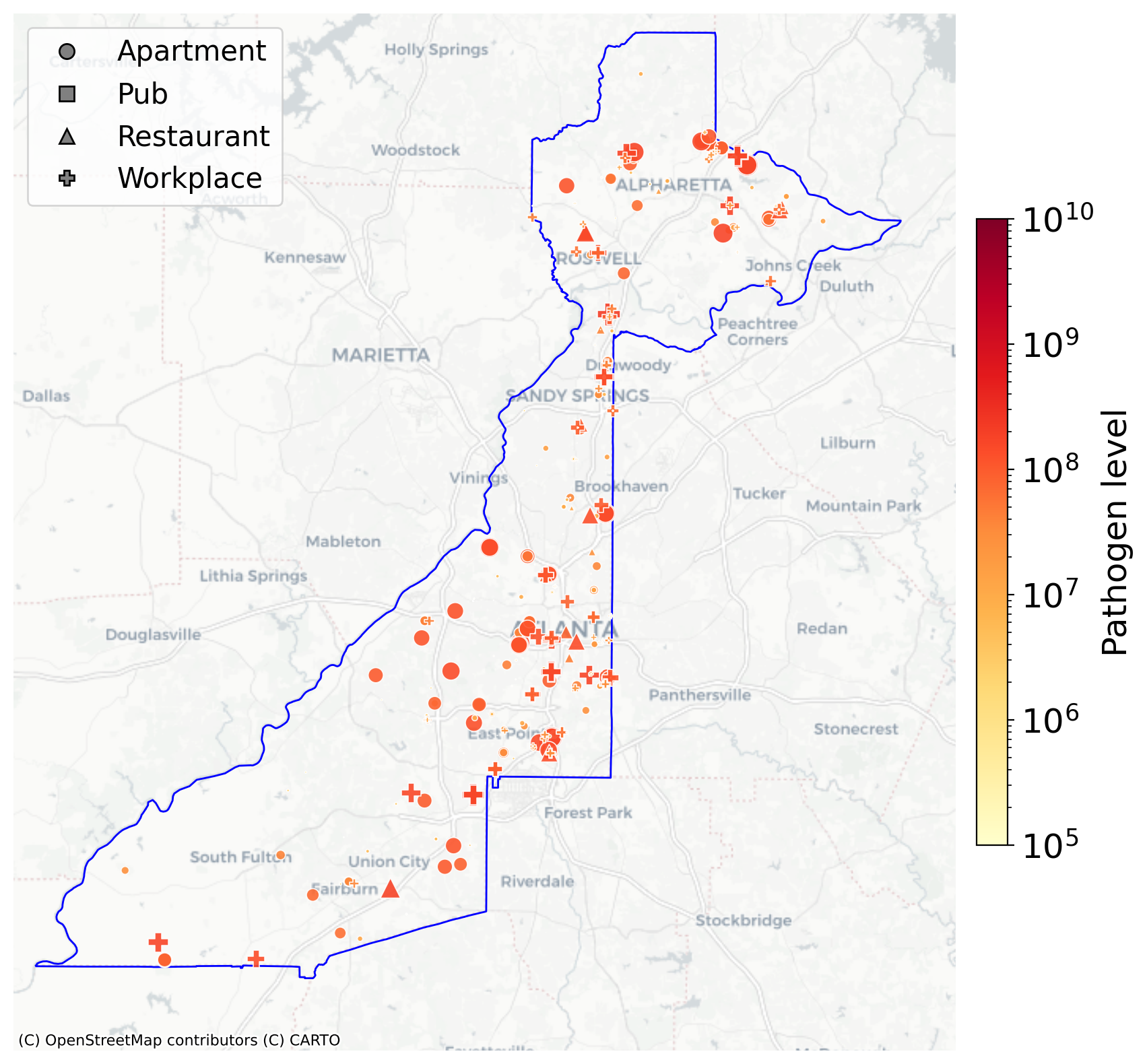}
        \caption{Highest transmission scenario (Infection rate = 0.5)}
        \label{subfig:spatial_feb1_ir_0_5}
    \end{subfigure}
    \vspace{-0.3cm}
    \caption{Spatial distribution of pathogen shedding in the Fulton County 10K simulation on February 1, 2024, across different infection rates. \fullresultsnote \vspace{-0.3cm}}\vspace{-0.3cm}
    \Description{Four maps of Fulton County showing the locations and magnitudes of pathogen shedding on February 1, 2024, for infection rates 0.1, 0.15, 0.25, and 0.5.}
    \label{fig:spatial_pathogen_fulton_infection_rates_feb_1}
\end{figure}

\begin{figure}[t]
    \centering
    \begin{subfigure}[b]{0.49\linewidth}
        \centering
        \includegraphics[width=\linewidth]{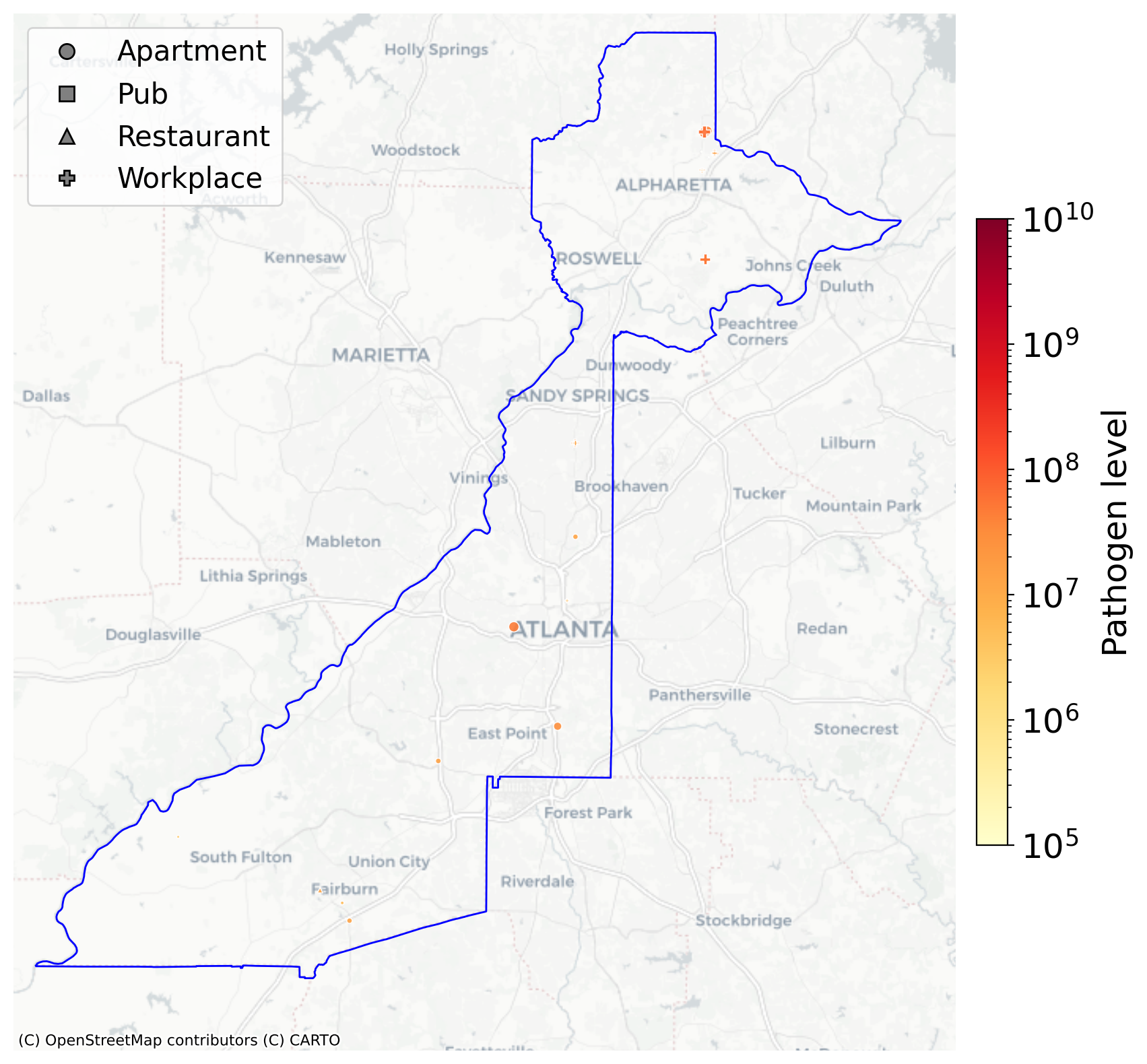}
        \caption{Lowest transmission scenario (Infection rate = 0.1)}
        \label{subfig:spatial_mar1_ir_0_1}
    \end{subfigure}
    \begin{subfigure}[b]{0.49\linewidth}
        \centering
        \includegraphics[width=\linewidth]{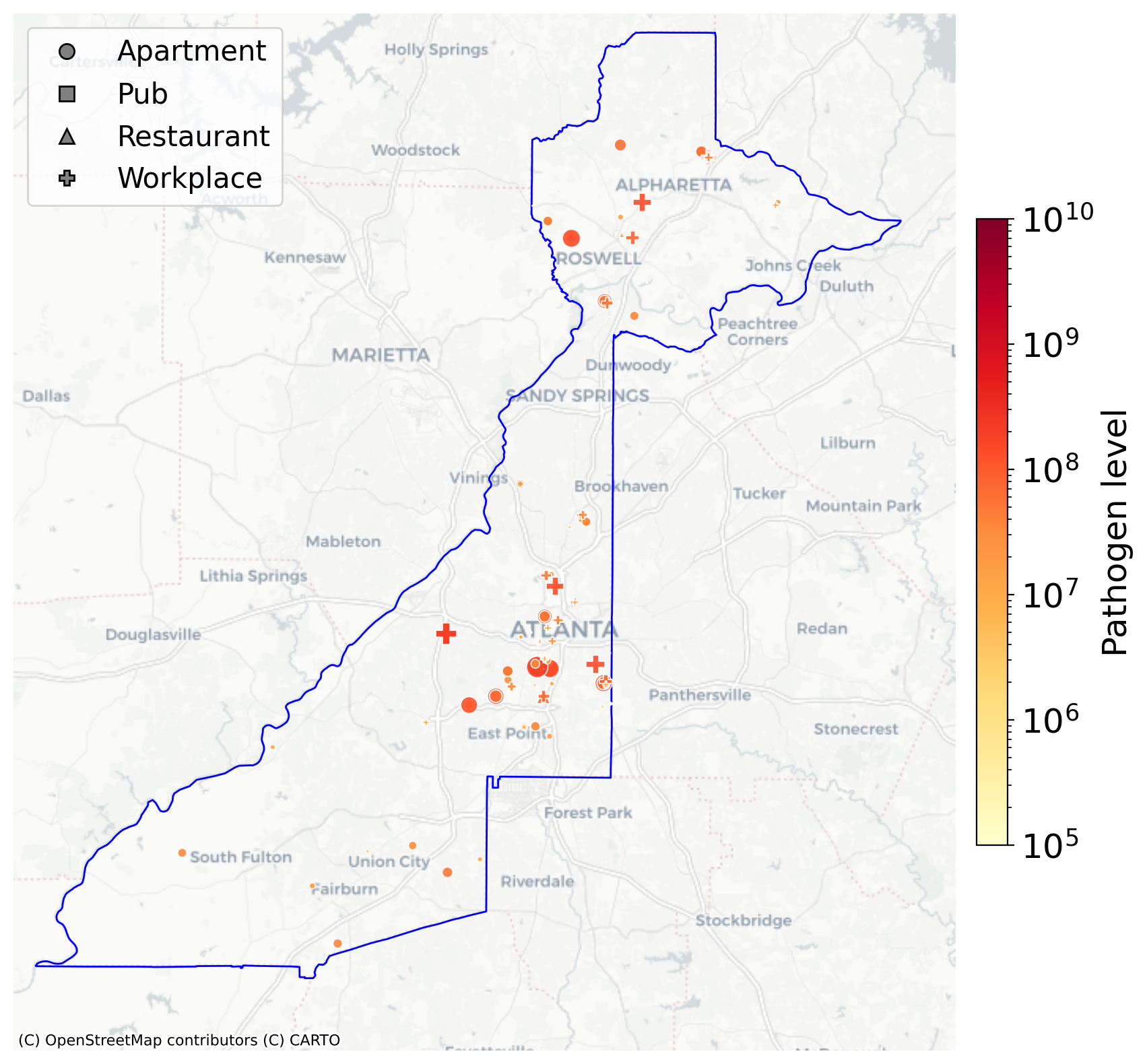}
        \caption{Slightly higher transmission scenario (Infection rate = 0.15)}
        \label{subfig:spatial_mar1_ir_0_15}
    \end{subfigure}
    \begin{subfigure}[b]{0.49\linewidth}
        \centering
        \includegraphics[width=\linewidth]{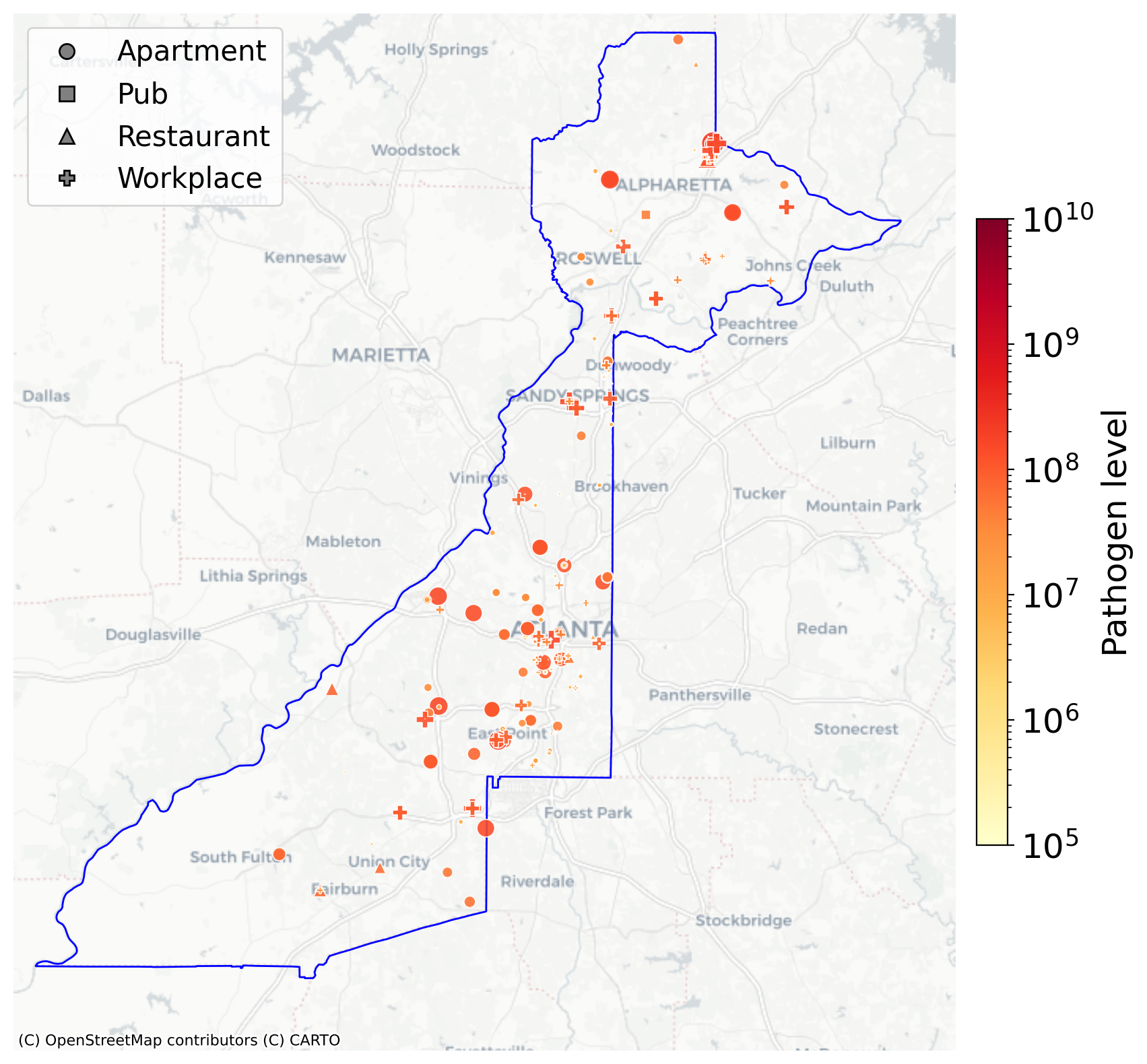}
        \caption{Moderate transmission scenario (Infection rate = 0.25)}
        \label{subfig:spatial_mar1_ir_0_25}
    \end{subfigure}
    \begin{subfigure}[b]{0.49\linewidth}
        \centering
        \includegraphics[width=\linewidth]{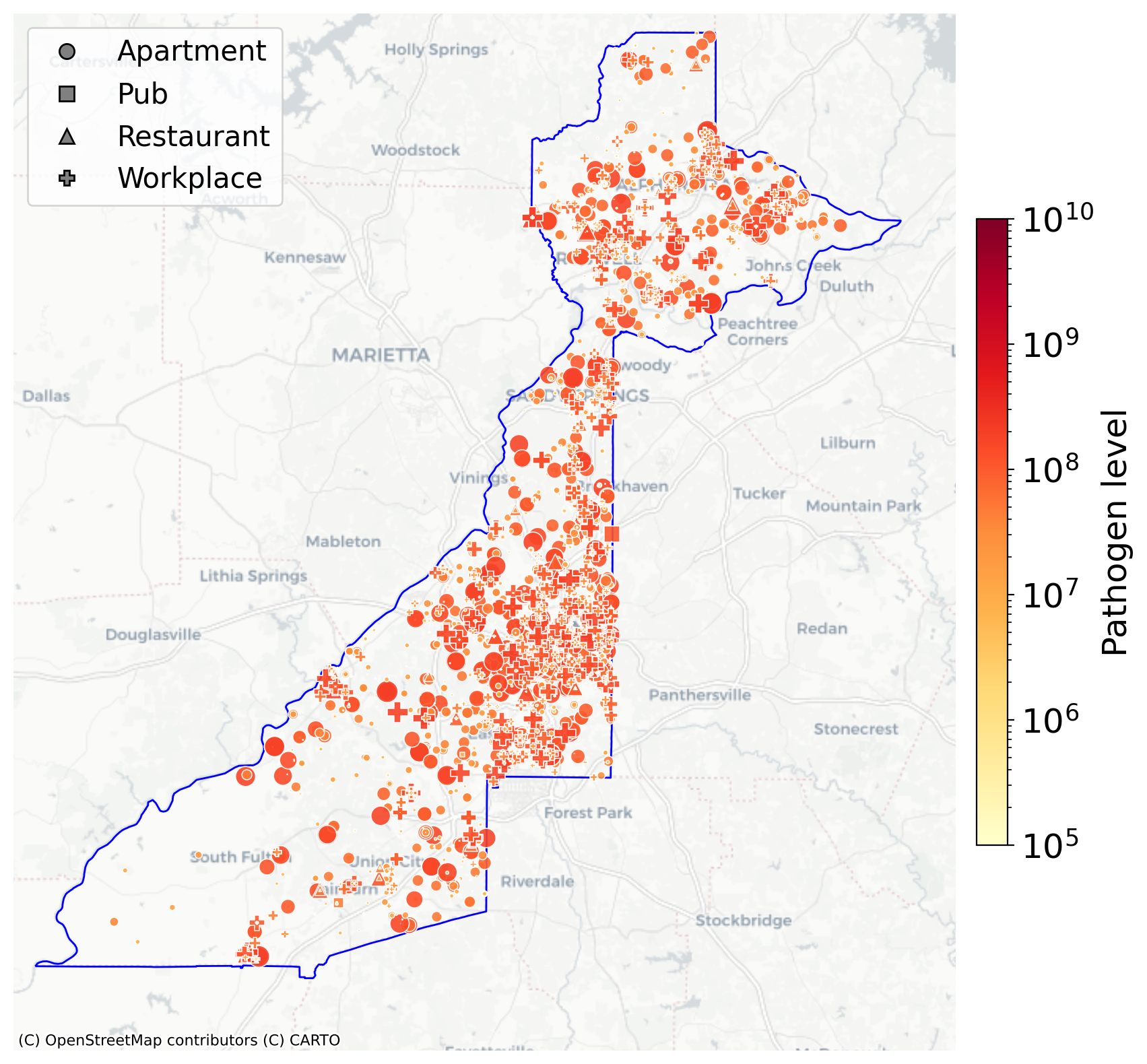}
        \caption{Highest transmission scenario (Infection rate = 0.5)}
        \label{subfig:spatial_mar1_ir_0_5}
    \end{subfigure}\vspace{-0.3cm}
    \caption{Spatial distribution of pathogen shedding in the Fulton County 10K simulation on March 1, 2024, across different infection rates. \fullresultsnote \vspace{-0.1cm}} \vspace{-0.1cm}
    \Description{Four maps of Fulton County showing the locations and magnitudes of pathogen shedding on March 1, 2024, for infection rates 0.1, 0.15, 0.25, and 0.5.}
    \label{fig:spatial_pathogen_fulton_infection_rates_mar_1}
\end{figure}

\begin{figure}[t]
    \centering
    \begin{subfigure}[b]{0.49\linewidth}
        \centering
        \includegraphics[width=\linewidth]{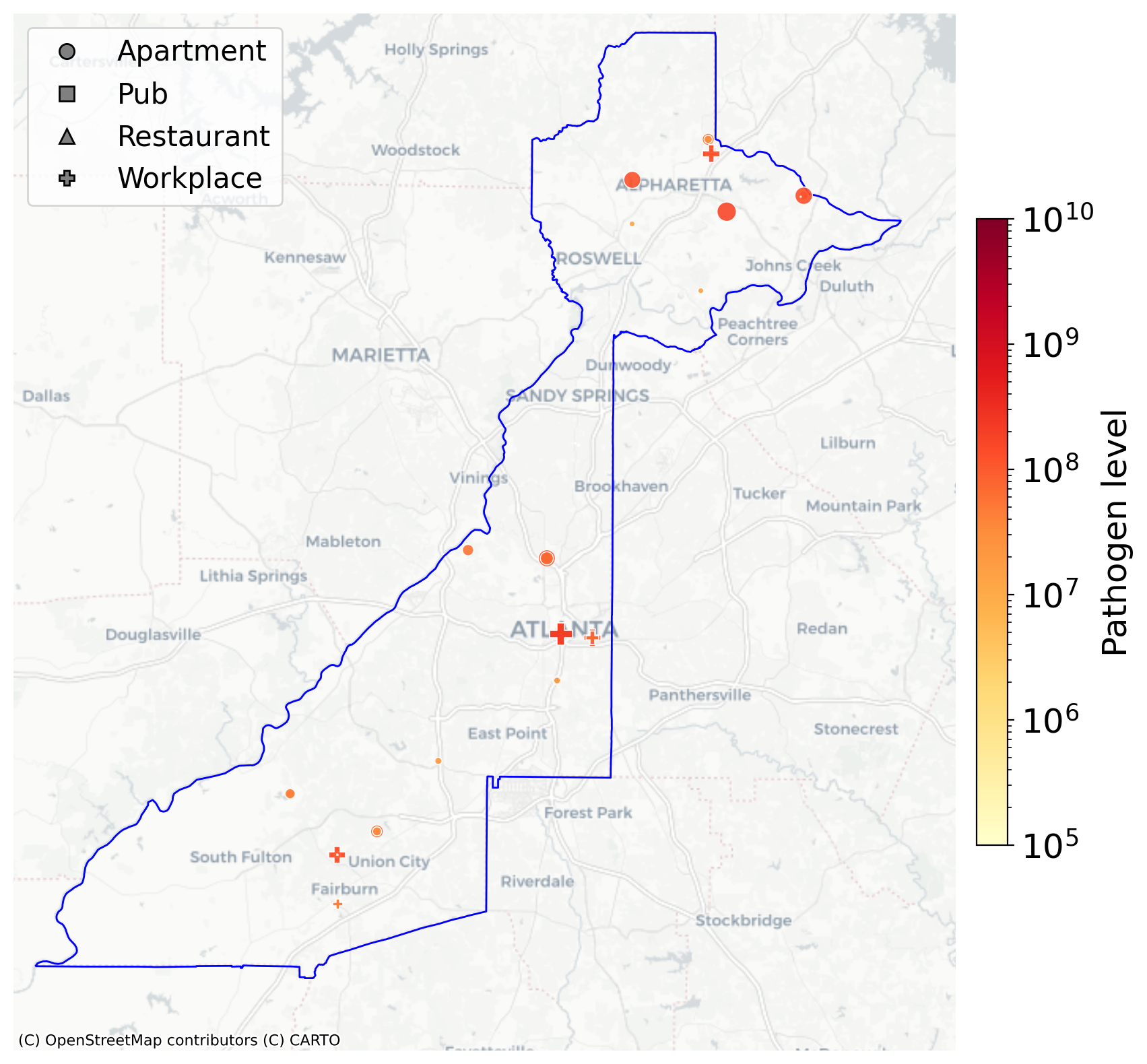}
        \caption{Lowest transmission scenario (Infection rate = 0.1)}
        \label{subfig:spatial_apr1_ir_0_1}
    \end{subfigure}
    \begin{subfigure}[b]{0.49\linewidth}
        \centering
        \includegraphics[width=\linewidth]{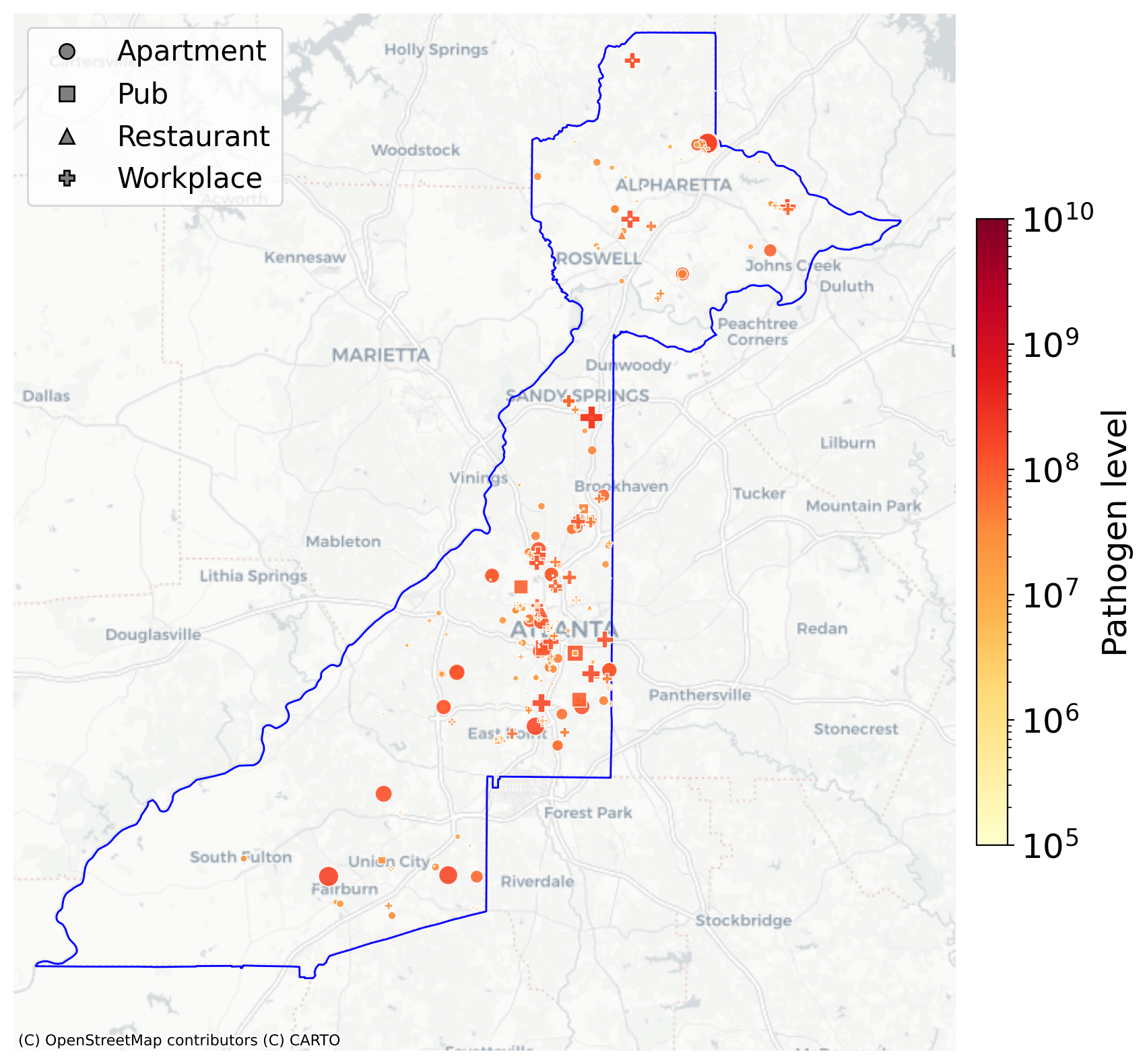}
        \caption{Slightly higher transmission scenario (Infection rate = 0.15)}
        \label{subfig:spatial_apr1_ir_0_15}
    \end{subfigure}
    \begin{subfigure}[b]{0.49\linewidth}
        \centering
        \includegraphics[width=\linewidth]{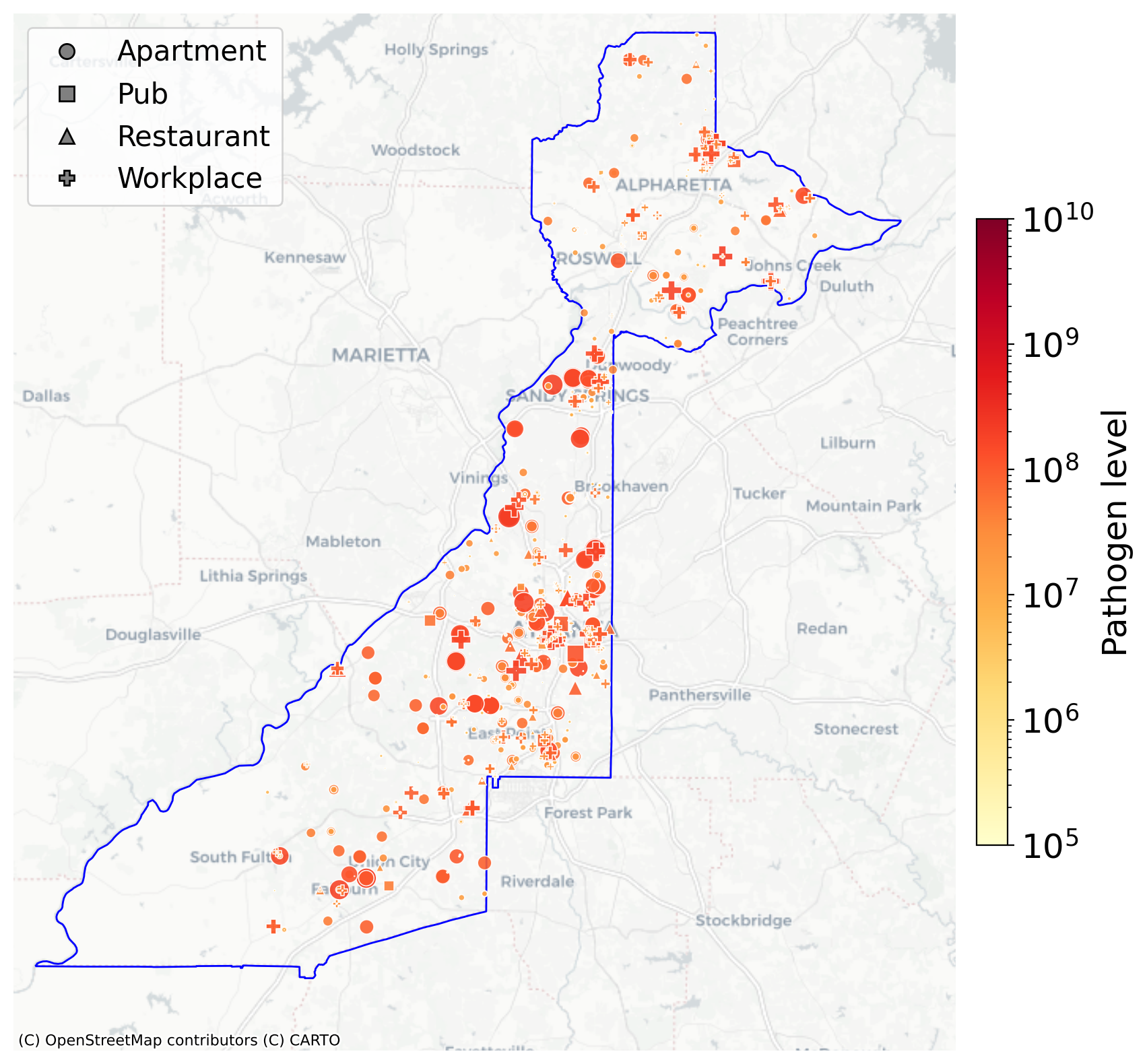}
        \caption{Moderate transmission scenario (Infection rate = 0.25)}
        \label{subfig:spatial_apr1_ir_0_25}
    \end{subfigure}
    \begin{subfigure}[b]{0.49\linewidth}
        \centering
        \includegraphics[width=\linewidth]{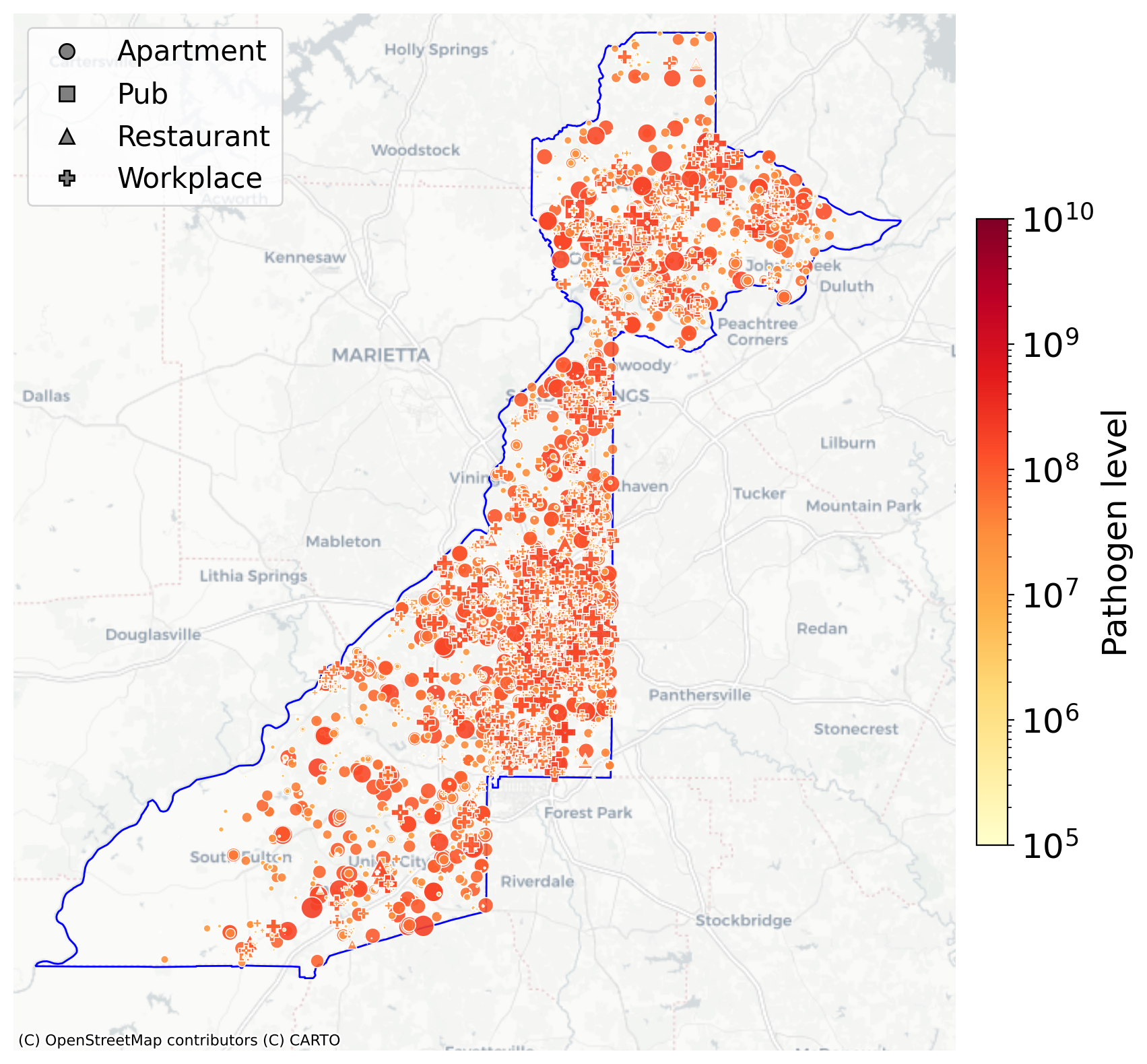}
        \caption{Highest transmission scenario (Infection rate = 0.5)}
        \label{subfig:spatial_apr1_ir_0_5}
    \end{subfigure}\vspace{-0.3cm}
    \caption{Spatial distribution of pathogen shedding in the Fulton County 10K simulation on April 1, 2024, across different infection rates. \fullresultsnote \vspace{-0.1cm}}\vspace{-0.1cm}
    \Description{Four maps of Fulton County showing the locations and magnitudes of pathogen shedding on April 1, 2024, for infection rates 0.1, 0.15, 0.25, and 0.5.}
    \label{fig:spatial_pathogen_fulton_infection_rates_apr_1}
\end{figure}

\begin{figure}[!ht]
    \centering
    \begin{subfigure}[b]{0.49\linewidth}
        \centering
        \includegraphics[width=\linewidth]{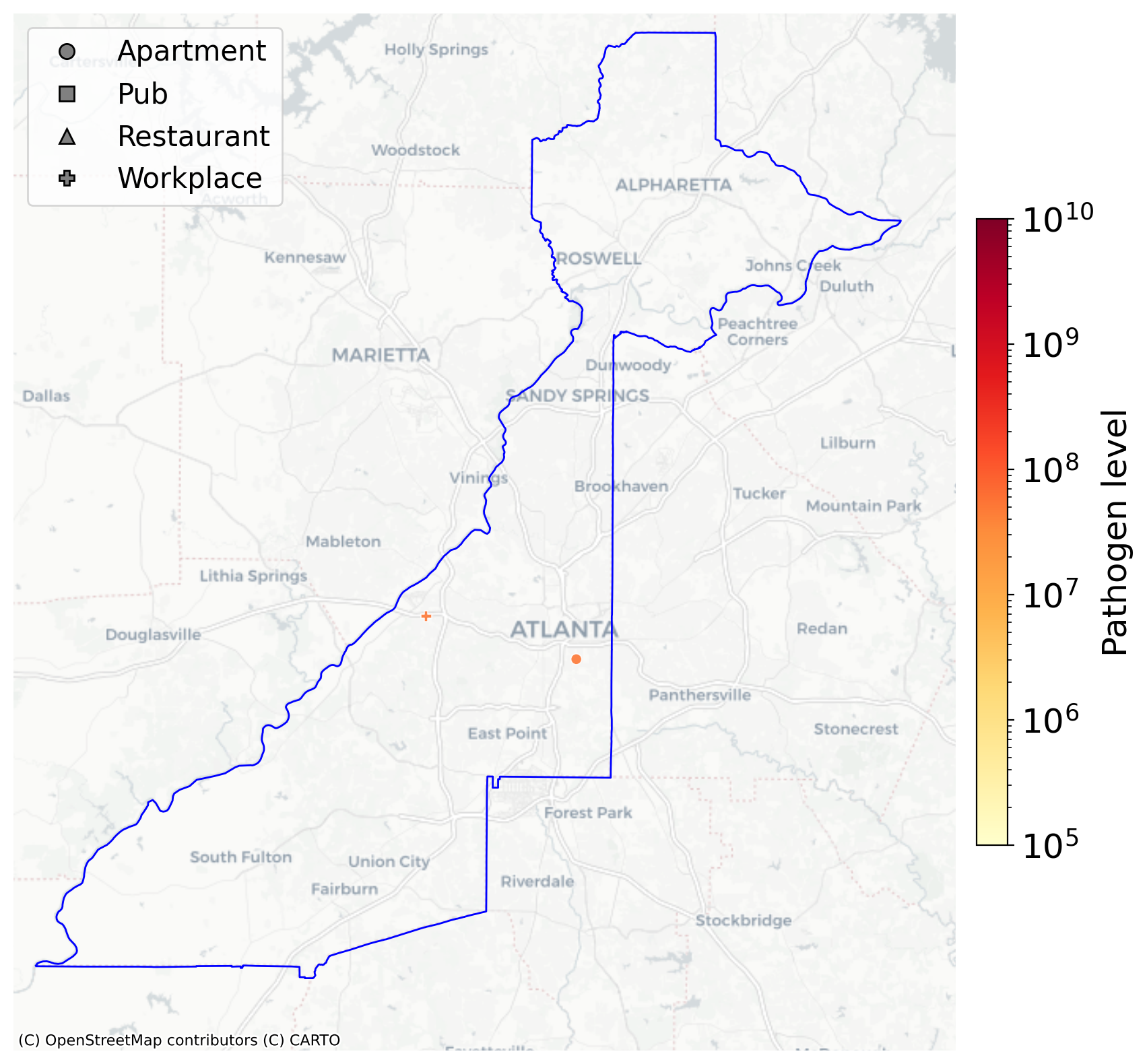}
        \caption{Lowest transmission scenario (Infection rate = 0.1)}
        \label{subfig:spatial_may1_ir_0_1}
    \end{subfigure}
    \begin{subfigure}[b]{0.49\linewidth}
        \centering
        \includegraphics[width=\linewidth]{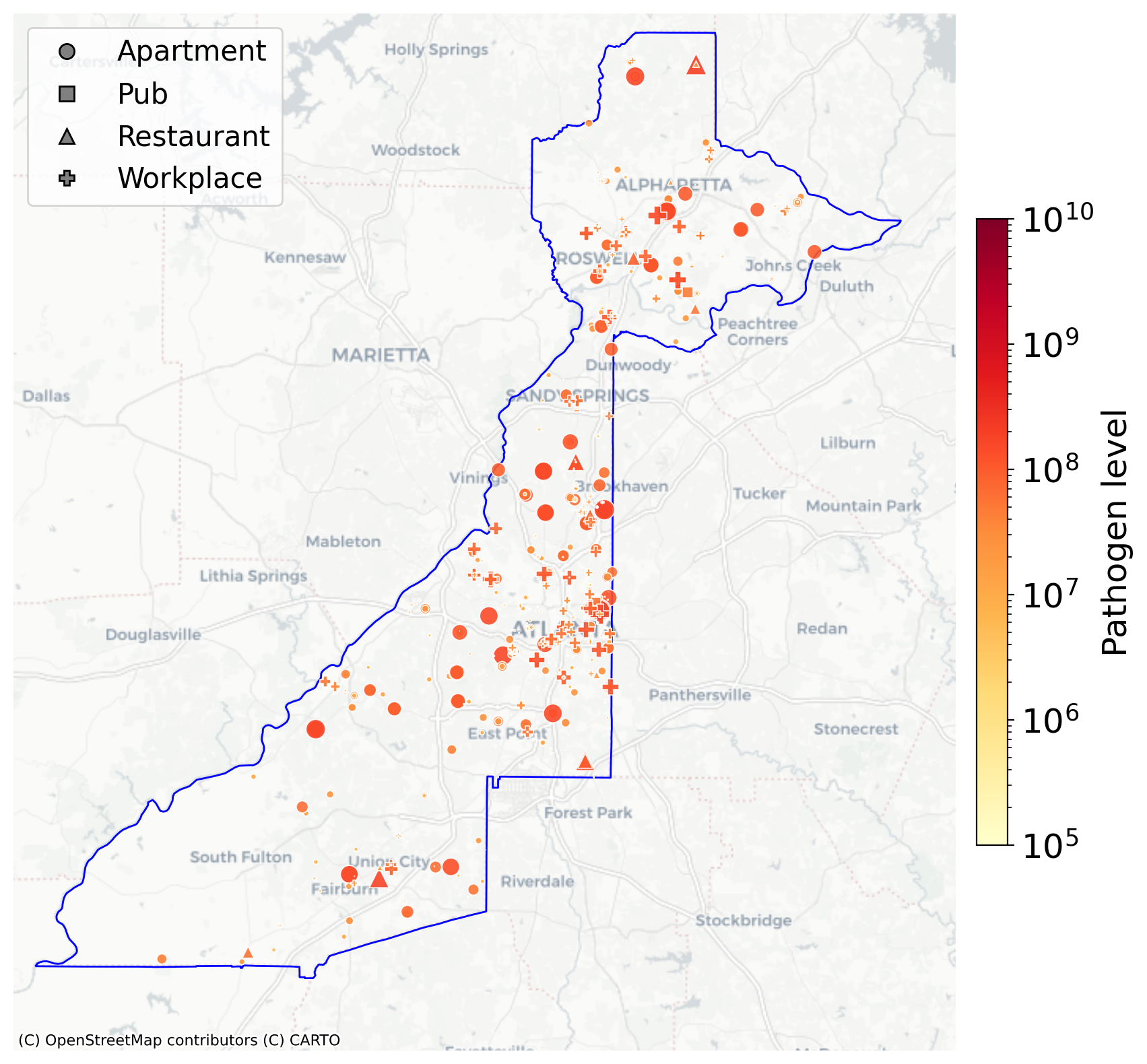}
        \caption{Slightly higher transmission scenario (Infection rate = 0.15)}
        \label{subfig:spatial_may1_ir_0_15}
    \end{subfigure}
    \begin{subfigure}[b]{0.49\linewidth}
        \centering
        \includegraphics[width=\linewidth]{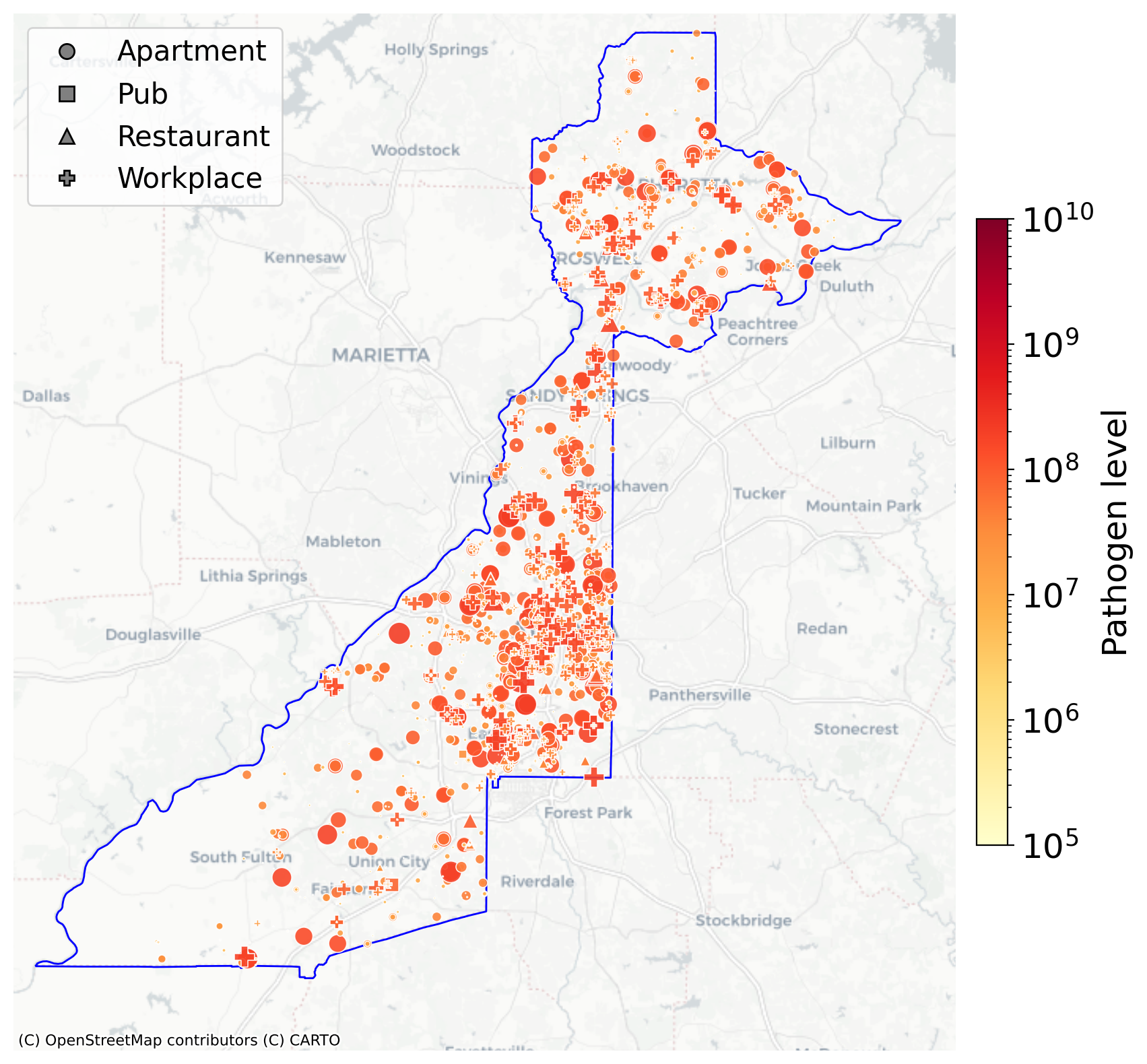}
        \caption{Moderate transmission scenario (Infection rate = 0.25)}
        \label{subfig:spatial_may1_ir_0_25}
    \end{subfigure}
    \begin{subfigure}[b]{0.49\linewidth}
        \centering
        \includegraphics[width=\linewidth]{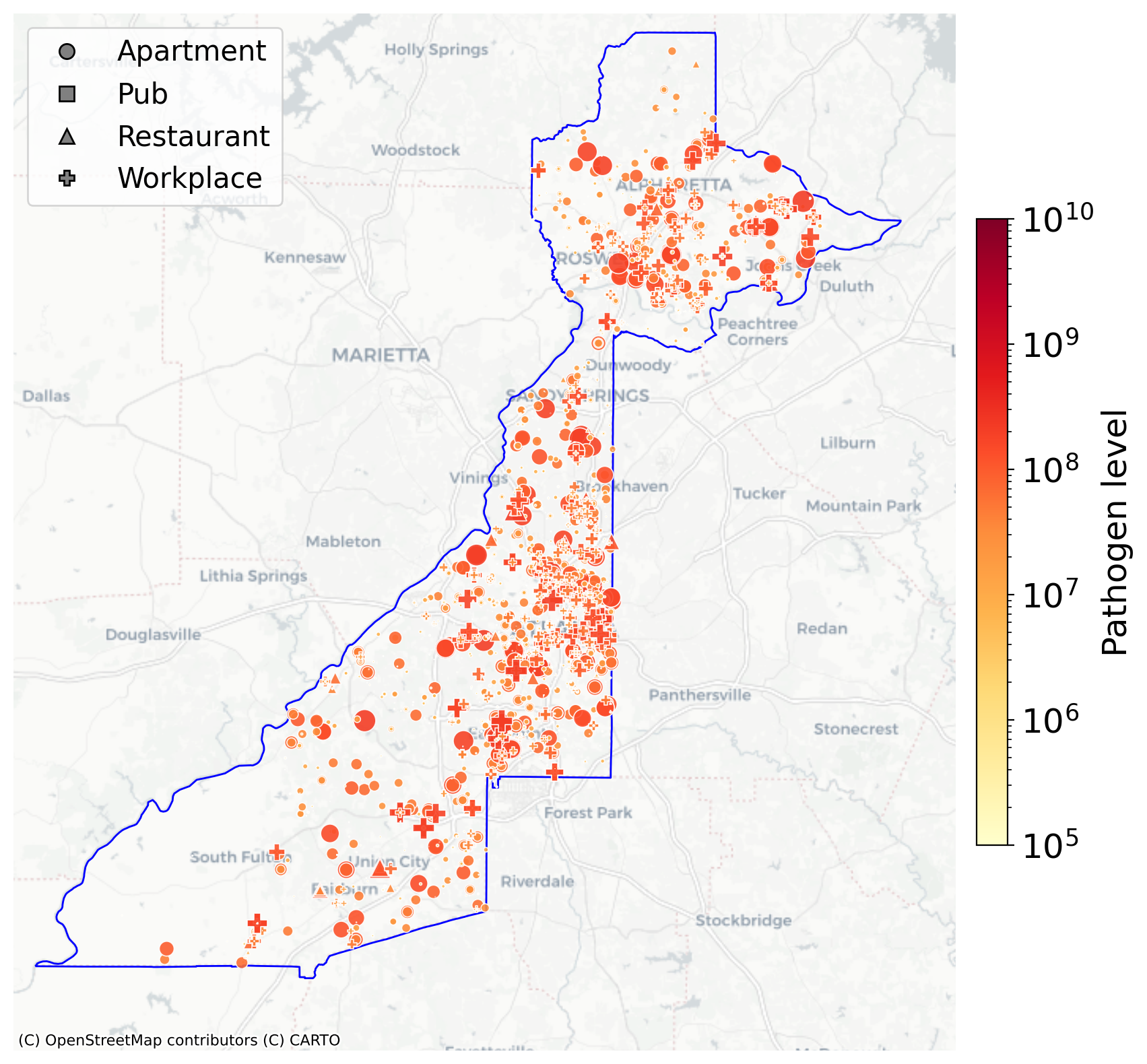}
        \caption{Highest transmission scenario (Infection rate = 0.5)}
        \label{subfig:spatial_may1_ir_0_5}
    \end{subfigure} \vspace{-0.3cm}
    \caption{Spatial distribution of pathogen shedding in the Fulton County 10K simulation on May 1, 2024, across different infection rates. \fullresultsnote \vspace{-0.3cm}}\vspace{-0.3cm}
    \Description{Four maps of Fulton County showing the locations and magnitudes of pathogen shedding on May 1, 2024, for infection rates 0.1, 0.15, 0.25, and 0.5.}
    \label{fig:spatial_pathogen_fulton_infection_rates_may_1}
\end{figure}

\reffig{fig:spatial_pathogen_fulton_infection_rates_feb_1},
\reffig{fig:spatial_pathogen_fulton_infection_rates_mar_1},
\reffig{fig:spatial_pathogen_fulton_infection_rates_apr_1}, and
\reffig{fig:spatial_pathogen_fulton_infection_rates_may_1}
illustrate the spatial progression of pathogen shedding in the Fulton County 10K simulation on
February~1, March~1, April~1, and May~1, 2024, respectively, across four infection rate
scenarios (0.1, 0.15, 0.25, and 0.5). The maps demonstrate that higher transmission
rates generate faster and more extensive spatial dissemination as the epidemic develops.
Greater infection potential accelerates both the establishment of new shedding sites
and their expansion across the county. As outbreaks intensify, the number of
locations receiving shed pathogen increases and pathogen loads rise, resulting in stronger,
denser, and more geographically connected spatial signatures over time. However, the highest infection rate triggers a much earlier and more intense outbreak, rapidly depleting the susceptible population. As a result, most individuals transition into the recovered state sooner, leaving fewer people available for new infections and greatly reducing pathogen contributions to the wastewater system in the later stages of the epidemic.

\reffig{fig:spatial_pathogen_fulton_infection_rates_feb_1} shows the spatial
distribution of early outbreak activity on February~1, 2024, under four different
infection rates. At this initial stage, pathogen shedding remains limited in all
scenarios, but clear differences emerge based on transmission intensity.
\reffig{subfig:spatial_feb1_ir_0_1}:
Only a few locations show low pathogen levels, indicating that transmission has not expanded significantly beyond initial introduction points. The magnitude of the pathogen load at each positive site appears high because it represents a per-agent shedding value defined in the simulation, which remains independent of the infection rate.
\reffig{subfig:spatial_feb1_ir_0_15}:
A few additional shedding sites appear, increasing the total number of affected locations compared with the lowest transmission scenario. Spatial spread remains constrained, but early signs of community transmission begin to emerge.
\reffig{subfig:spatial_feb1_ir_0_25}:
Multiple clusters are visible across northern and central Fulton County, indicating
stronger early propagation. Higher shedding intensity suggests that the outbreak is
transitioning into sustained exponential growth.
\reffig{subfig:spatial_feb1_ir_0_5}:
Large numbers of sites already exhibit substantial pathogen loads. Rapid, widespread
seeding of infections has taken place across major residential and commercial areas,
highlighting an aggressive early expansion of the epidemic.

\reffig{fig:spatial_pathogen_fulton_infection_rates_mar_1} shows the spatial footprint
of pathogen shedding one month later, on March~1, 2024. By this point, transmission
differences between infection rate scenarios have become more pronounced, with higher
rates producing broader spread and higher pathogen loads across the county.
\reffig{subfig:spatial_mar1_ir_0_1}:
Only a few scattered shedding sites remain detectable, all at low levels. Spatial
propagation has largely stalled, indicating minimal community spread and a contained
outbreak trajectory.
\reffig{subfig:spatial_mar1_ir_0_15}:
Clusters of pathogen shedding emerge in central neighborhoods and along major mobility
corridors. Although shedding remains moderate, infections are now persisting and
expanding within connected social environments.
\reffig{subfig:spatial_mar1_ir_0_25}:
Shedding becomes widespread, forming several high-intensity hotspots. The pathogen
spreads efficiently across residential, commercial, and workplace settings, reflecting
a rapidly growing epidemic with strong spatial connectivity.
\reffig{subfig:spatial_mar1_ir_0_5}:
Nearly all populated regions exhibit high pathogen loads with dense clusters. This indicates extensive and mature community
transmission, with few remaining unaffected areas across the county.

\reffig{fig:spatial_pathogen_fulton_infection_rates_apr_1} shows the spatial
distribution of pathogen shedding on April~1, 2024, when transmission has progressed
farther and spatial differences between scenarios are clearly established. Higher
infection rates produce widespread clusters of pathogen shedding, while lower rates remain
largely localized and constrained.
\reffig{subfig:spatial_apr1_ir_0_1}:
Only a few low-intensity detections persist. Transmission remains highly limited, with
no indication of significant expansion or sustained pathogen load in wastewater.
\reffig{subfig:spatial_apr1_ir_0_15}:
Shedding becomes more spatially organized, forming multiple clusters around Atlanta and
northern communities. Community spread is evident, though still moderate in scale.
\reffig{subfig:spatial_apr1_ir_0_25}:
An area with a large and continuous pathogen load emerges along the major population corridor.
Many locations show strong shedding signals, demonstrating that the epidemic is now
well established throughout the county.
\reffig{subfig:spatial_apr1_ir_0_5}:
Widespread, high pathogen loads cover nearly the entire region. High-shedding
sites dominate both urban centers and suburban zones, indicating a fully developed
outbreak with pervasive community transmission.

\reffig{fig:spatial_pathogen_fulton_infection_rates_may_1} shows the spatial extent of
pathogen shedding on May~1, 2024. By this time, the epidemic has either remained
localized or expanded dramatically depending on the infection rate. The resulting
spatial patterns highlight the long-term consequences of different transmission
intensities.
\reffig{subfig:spatial_may1_ir_0_1}:
Only a couple of isolated, low-intensity detections remain. The outbreak never achieves
sustained transmission and is effectively extinguished.
\reffig{subfig:spatial_may1_ir_0_15}:
Shedding spreads further within clustered regions, primarily in central and northern parts of the county. Transmission continues, but the rate of spatial expansion remains relatively low.
\reffig{subfig:spatial_may1_ir_0_25}:
A large fraction of facilities show strong pathogen loads, reflecting widespread and
ongoing community transmission supported by dense spatial connectivity.
\reffig{subfig:spatial_may1_ir_0_5}:
Shedding begins to decline as the susceptible population becomes exhausted and infected agents transition into the recovered state, where they can no longer be reinfected or contribute additional pathogen to the system.

Since it is difficult to convey infectious disease dynamics and shedding events using static screenshots, we have created a dynamic visualization found at \url{https://poop-simcity.pages.dev/}. This demonstration shows the movement of simulated agents in Fulton County over time, highlights exposed and infectious agents, illustrates infectious shedding events, and shows the shed pathogen levels at each simulation time.

\subsubsection{Effect of Human Mobility on Wastewater-Based Epidemiology}

\begin{figure}
    \centering
    \includegraphics[width=0.9\linewidth, height=3in]{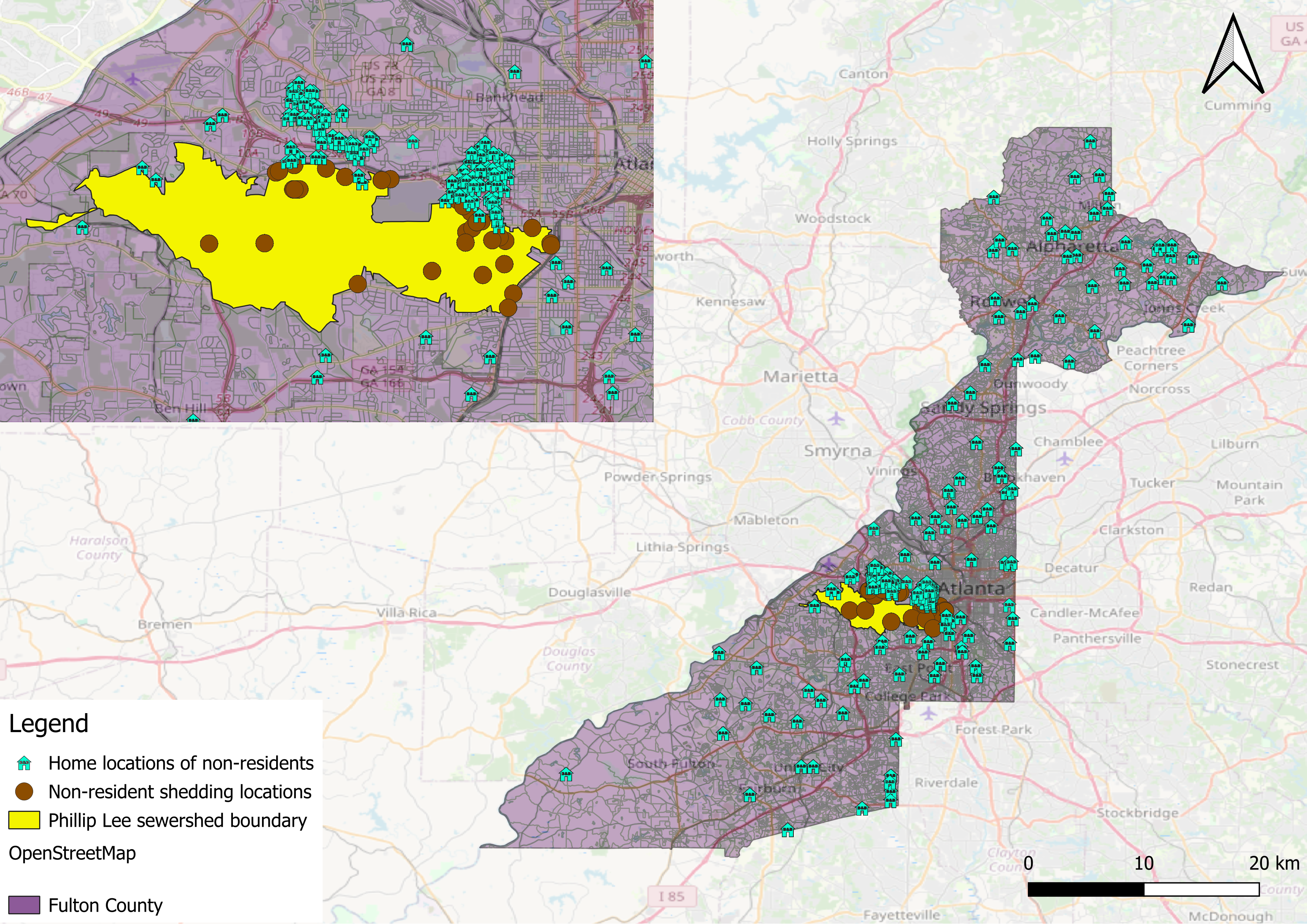}
    \caption{Shedding locations within Phillip Lee sewershed and home locations of nonresident shedding agents across Fulton County.}
    \Description{Map highlighting the Phillip Lee sewershed, shedding locations inside it, and the home locations of nonresident agents who shed within the sewershed.}
    \label{fig:p1}
\end{figure}

\begin{figure}
    \centering
    \includegraphics[width=0.9\linewidth, height=3in]{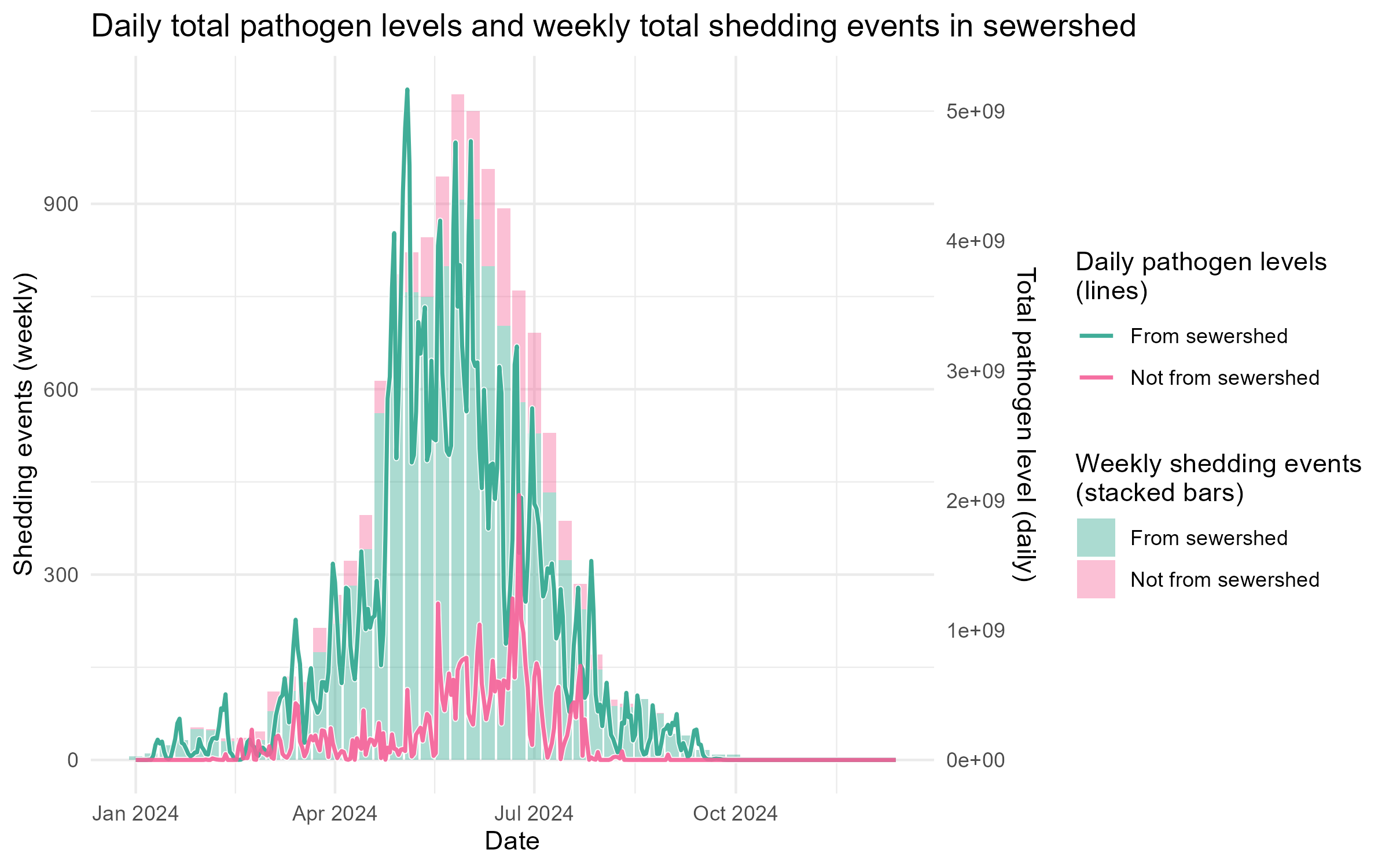}
    \caption{Daily shedding events and pathogen loads contributed by infectious individuals within the Phillip Lee sewershed.}
    \Description{Time series comparing daily shedding-event counts and pathogen loads from sewershed residents and nonresidents during the simulation.}
    \label{fig:p2}
\end{figure}

% \begin{figure}[htbp]
%     \centering
%     \begin{subfigure}[b]{0.49\linewidth}
%         \centering
%         \includegraphics[width=0.9\linewidth]{figs/aksh1.png}
%         \caption{Daily total pathogen levels resulting from shedding events within the sewershed.}
%         \label{fig:p1}
%     \end{subfigure}
%     \hfill
%     \begin{subfigure}[b]{0.49\linewidth}
%         \centering
%         \includegraphics[width=0.9\linewidth]{figs/aksh2.png}
%         \caption{Daily shedding events from infectious individuals within the sewershed.}
%         \label{fig:p2}
%     \end{subfigure}

%     \caption{Simulated pathogen shedding dynamics within the sewershed: (a) daily total pathogen levels and (b) the number of shedding events generated by infectious individuals.}
%     \label{fig:pathogen-shedding}
% \end{figure}

To understand how the spatial distribution of shedding events affects wastewater-based epidemiology, this section uses a specific sewershed within Fulton County as a case study. \reffig{fig:p1} shows the location of the Phillip Lee sub-sewershed within Fulton County. We chose this sewershed as it covers a highly populated area in downtown Atlanta and covers both industrial and municipal infrastructure zones. In addition to the spatial boundaries of this sewershed, \reffig{fig:p1} also shows the home location of agents not living within the Phillip Lee sewershed boundaries (non-residents) who shedded (defected) with the boundaries and the exact locations within the Phillip Lee sewershed where the agents shedded. We see that many non-resident shedders came from nearby regions, but we also see some non-residents who travelled from across the Fulton County simulation area. 

To understand the impact of this mobility on wastewater-based epidemiology, \reffig{fig:p2} shows the number of shedding events and the daily pathogen load contributed by infectious individuals within the sewershed for the scenario with an infection rate of 0.25. The number of shedding events follows the number of infectious individuals in the corresponding scenario shown in \reffig{subfig:fulton_ir_0_25}: both increase during the first half of the simulation, peak around June, and decline thereafter. For both sewershed residents and nonresidents, shedding events occur more often on weekdays than on weekends, producing the recurring spikes in the figure. Although the two groups follow similar overall trends, their peaks differ. Resident shedding events peak in June 2024, whereas the nonresident peak extends from June into July. The daily pathogen loads show a similar midyear pattern, with the resident contribution peaking in June and the nonresident contribution peaking in July.

\section{Discussions}
\label{sec:discussions}
A major challenge in leveraging WBE for public health decision making lies in accurately defining the representativeness of WWS. This includes determining the proportion of infections detected (numerator) relative to the size of the population captured (denominator) and understanding the spatial distribution of the contributing population. Accurate catchment information is critical for normalizing wastewater measurements (i.e., transforming wastewater concentration to metrics that can be more comparable between samples), enabling timely interventions, and pinpointing target areas for public health action. While the WWS at downstream wastewater treatment facilities predominantly relies on sewer networks to define its
catchment, the WWS catchment at the upstream site was more impacted by human behavior and social networks.

Our current model assumes that pathogen shed at a toilet reaches the sampling site instantaneously and without loss. In reality, two in-sewer processes shape the temporal relationship between pathogen shedding and detection/measurement \cite{wang2020designing}. First, hydraulic transport imposes a travel time between a shedding event location and its arrival at the sampling location that varies with distance, flow velocity, and network length, ranging from minutes for nearby shedding events to several hours for distant ones, while longitudinal dispersion could spread a sharp shedding pulse over time, broadening and flattening the measured peak. Second, genetic biomarkers of pathogens decay in the sewer through temperature-dependent degradation, microbial activity, and sorption, commonly approximated as first-order decay in residence time \cite{ahmed2020decay, fu2023longitudinal}; this disproportionately suppresses contributions of shedding events that travel farther or reside longer in the network (e.g., during low-flow periods), and biases the apparent spatial distribution toward sources near the sampling location. Because both effects scale with travel time, they are largest for big downstream WWTP sites and smallest for upstream sites close to their shedding event locations. Incorporating them could therefore delay detection relative to shedding, lower and broaden the observed peak, and reduce the total measured load, so our present results represent an upper bound on the speed and clarity of the wastewater pathogen load. Because we report daily aggregated loads over a multi-month epidemic, we expect these processes to shift and smooth the peak rather than alter the qualitative correspondence between wastewater loads and epidemic dynamics; nevertheless, they would be important for sub-daily interpretation and for comparing loads across sampling locations. These processes can be integrated in future work by coupling the individual-level shedding output with a hydraulic and water-quality sewer model that propagates loads with realistic flow and a residence-time- and temperature-dependent decay term.

\section{Conclusions}
\label{sec:conclusion}
In this work, we presented an agent-based geospatial simulation framework that links human behavior, infectious disease dynamics, and wastewater pathogen loads for WBE. Building on the Patterns of Life simulation, we integrated daily mobility, multilayer social interactions, physiologically motivated defecation, and an individual-level SEIR process with gamma-like shedding dynamics. Using a case study of 10,000 agents in Fulton County, Georgia, we demonstrated how infection rates, mobility patterns, and toilet use jointly shape epidemic trajectories, wastewater pathogen loads, and the spatial distribution of pathogen shedding.

Our results highlight two insights that rely on the individual-level, behavior- and mobility-explicit nature of the framework. First, mobility and toilet use can substantially decouple residential case counts from wastewater contributions, especially at upstream sampling locations, explaining why nominal sewershed populations can produce different wastewater pathogen loads. Second, spatial analyses reveal how shedding clusters emerge, grow, and dissipate over time, revealing potential transmission corridors and hotspots that are not visible in aggregated wastewater measurements alone. As a consistency check, the aggregate wastewater load closely tracks epidemic timing and magnitude and, as expected for SEIR-type dynamics, responds nonlinearly to the infection rate because early depletion of susceptible agents limits transmission in high-transmission scenarios.

This work also has limitations that motivate future research. The current implementation models a single pathogen using a simplified SEIR structure and does not include hydraulic transport or in-sewer decay. Future extensions will incorporate realistic sewer flow, multiple pathogens, and behavioral changes in response to interventions. Calibration against empirical wastewater and case data will further refine model parameters and improve interpretability.

Overall, our findings show that behaviorally grounded agent-based simulations can serve as a virtual laboratory for interpreting wastewater pathogen loads, evaluating surveillance-system design, and exploring intervention scenarios. By explicitly connecting human movement, toilet use, and within-host shedding to wastewater measurements, this framework supports improved design and analysis of wastewater-based epidemiology in complex and dynamic urban environments.

\section{Acknowledgment}
\label{sec:acknowledgment}
This publication was made possible by the Insight Net cooperative agreement CDC-RFA-FT-23-0069 from CDC’s Center for Forecasting and Outbreak Analytics. Its contents are solely the responsibility of the authors and do not necessarily represent the official views of the Centers for Disease Control and Prevention.
 This research was also supported by NSF award \#2109647.
  
We also acknowledge the use of ChatGPT 5.x to assist with English-language editing of the manuscript.

\bibliographystyle{ACM-Reference-Format}
\bibliography{main}

\end{document}